\documentclass[aps,prb,twocolumn,superscriptaddress,
groupedaddress,10pt]{revtex4}
\usepackage{amsmath}
\usepackage{amssymb}
\usepackage{graphicx}
\usepackage{epsfig}
\usepackage{dcolumn}
\usepackage{bm}
\usepackage{blindtext}
\usepackage{natbib}
\usepackage{siunitx} 
\usepackage{multirow}
\usepackage{bbm}

\usepackage{empheq,accents}
\usepackage{amsthm}

\newtheorem{Theorem}{Theorem}

\newtheorem{Lemma}{Lemma}

\newtheorem{Prop}{Proposition} 

\usepackage[usenames,dvipsnames]{xcolor}
\usepackage{amsthm}
\usepackage{booktabs}
\usepackage{hyperref}
\usepackage{tikz}
\usetikzlibrary{calc}
\usepackage{mathtools}

\def\be{\begin{equation}} \def\ee{\end{equation}}
\def\bea{\begin{eqnarray}} \def\eea{\end{eqnarray}}

\def\nn{\nonumber}

\newcommand{\ket}[1]{| #1 \rangle}
\newcommand{\bra}[1]{\langle #1 |}

\begin{document}
\title{
Measurement-based quantum computation in symmetry protected topological states of one-dimensional  integer spin systems
}

\author{Wang Yang}
\affiliation{School of Physics, Nankai University, Tianjin, China}

\author{Arnab Adhikary}
\affiliation{Department of Physics and Astronomy, University of British Columbia, Vancouver, Canada}
\affiliation{Stewart Blusson Quantum Matter Institute, University of British Columbia, Vancouver, Canada}
\affiliation{Leibniz University Hannover, Hannover, Germany}

\author{Robert Raussendorf}
\affiliation{Leibniz University Hannover, Hannover, Germany}
\affiliation{Stewart Blusson Quantum Matter Institute, University of British Columbia, Vancouver, Canada}
%\orcid{0000-0003-4983-9213}

\begin{abstract}

In this work, we generalize the algebraic framework for measurement-based quantum computation (MBQC) in one-dimensional symmetry protected topological states recently developed in [Quantum 7, 1215 (2023)],
such that in addition to half-odd-integer spins, the integer spin chains can also be incorporated in the framework. 
The computational order parameter characterizing  the efficiency of MBQC is identified,
which, for spin-$1$ chains in the Haldane phase, coincides with the conventional string order parameter in condensed matter physics.

\end{abstract}

%\pacs{75.10.Pq, 05.30.Rt, 71.10.Hf, 75.10.Jm}
\maketitle

%%%%%%%%%%%%%%%%%%%%%%%%%%%%%%%%%%%%%%%%%%%%%%%%%%%%%%
\section{Introduction}

In condensed matter physics, symmetry protected topological (SPT) phases are non-symmetry broken gapped phases which are short-range entangled and cannot be adiabatically connected to product states without breaking the symmetries \cite{Gu2009,Schuch2011,Chen2011,Pollmann2012,Chen2013,Gu2014}. 
In one dimension, a typical SPT state is the Affleck-Kennedy-Lieb-Tasaki (AKLT) state in spin-$1$ chain \cite{Affleck1987,Affleck1988}, which belongs to the SPT nontrivial Haldane phase in integer spin chains protected by the $\mathbb{Z}_2\times \mathbb{Z}_2$ symmetry \cite{Haldane1983,Haldane1983b,Chen2011b}.
In the Haldane phase, the SPT order is characterized by the string order parameter \cite{Hida1992,Oshikawa1992,Garcia2008}:
The system is SPT non-trivial (trivial) if the string order parameter is non-vanishing (vanishing) in the long distance limit. 

In the past decade, it has been established that the available resource states for measurement-based quantum computation (MBQC) \cite{Raussendorf2001,Raussendorf2003} are not restricted to special many-body entangled states \cite{Gross2007a,Gross2007,Brennen2008}, 
but can actually be chosen from a multitude of states which constitute a phase, dubbed computational phase \cite{Doherty2009,Chung2009,Miyake2010,Else2012,Wei2012}. 
Interestingly, computational  phases coincide with SPT phases \cite{Raussendorf2017,Stephen2017,Raussendorf2019,Stephen2019}, revealing a deep connection between quantum computation and condensed matter physics. 
We will abbreviate MBQC in SPT phases as SPT-MBQC for short in this article. 

Although universal MBQC on resource states with bounded range of entanglement can only be achieved in spatial dimension two or higher \cite{Raussendorf2001,Raussendorf2006} (see Ref. \onlinecite{Stephen2024}, however),
one-dimension (1D) serves as a useful prototypical model for illustrating the MBQC methods,
in which a single logical qubit can be simulated. 
For the notions and classifications of computational phases in SPT-MBQC, 1D systems are the most well-studied up to date \cite{Else2012,Raussendorf2017,Stephen2017}.
Early discussions of 1D computationally nontrivial states were based on the language of matrix product states (MPS),
which requires the system size to be in the infinite limit. 
In a recent work Ref. \onlinecite{Raussendorf2023}, an algebraic approach to 1D SPT-MBQC  has been developed which applies to finite systems as well. 
%The discussions in Ref. \onlinecite{Raussendorf2023} are based on a number of fundamental assumptions (or ``axioms") for site-local or block-local linear and projective representations of the symmetry group. 
A key conclusion of Ref. \onlinecite{Raussendorf2023} is that as long as the string order parameter is nonzero, MQBC is able to simulate any single-qubit logical gate to any desired accuracy.
Therefore, string order parameters play the role as computational order parameters for characterizing the computational phases,
similar to their roles in 1D SPT phases, albeit now even applicable for finite systems. 
In addition, the efficiency of performing MBQC is determined by the string order parameter: 
The smaller the string order parameter, the larger the operational overhead.

On the other hand, although integer spin chains in the Haldane phase belong to the same SPT phase as the spin-$1/2$ cluster and dimerized XXZ chains, 
integer spin systems are beyond the scope of the algebraic formalism in Ref. \onlinecite{Raussendorf2023}.
The technical reason is that the existences of block-local projective representations in the bulk of the system are required for the formalism in Ref. \onlinecite{Raussendorf2023} to function,
whereas integer spin chains do not support physical projective representations on the sites. 
Hence, a modification of the MBQC  constituents in Ref. \onlinecite{Raussendorf2023} is needed to incorporate the cases of integer spins. 

In this work, we find that by a modification of the assumed representations, 
the algebraic SPT-MBQC formalism can be extended to  finite integer spin chains as well. 
The modified formalism works for any 1D qudit system, regardless of its spin value being half-odd-integer or integer. 
In particular, the corresponding computational order parameter in the modified formalism is identified, which, for the spin-$1$ chains in the Haldane phase, coincides with the conventional string order parameter in condensed matter physics. 

The rest of the paper is organized as follows. 
In Sec. \ref{sec:review}, we give a pedagogical review of MBQC in spin-$1$ AKLT chain. 
In Sec. \ref{sec:sum_paths}, a summing-over-all-measurement-path approach is discussed for SPT-MBQC in the spin-$1$ Haldane phase for the special and simple case of a single logical rotation
before going to the algebraic approach.
%which motivates the way for modifying the assumptions in Ref. \onlinecite{Raussendorf2023}.
In Sec. \ref{sec:algebraic_MBQC}, the modified algebraic formalism is discussed,
containing a detailed proof of the functioning of the formalism.
Sec. \ref{sec:applications} includes two examples for applications of the modified  formalism,
including the spin-$1$ Haldane phase and the formalism in Ref. \onlinecite{Raussendorf2023}. 
Finally, in Sec. \ref{sec:summary}, we briefly summarize the main results of this work.

%%%%%%%%%%%%%%%%%%%%%%%%%%%%%%%%%%%%%%%%%%%%%%%%%%%%%%
\section{Review of MBQC in 1D spin-$1$ AKLT chain}
\label{sec:review}

In this section, we review the spin-$1$ AKLT chain and how to perform MBQC in it.

%------------------------------------------------------------------------------------------------------------------------------
\subsection{Spin-$1$ AKLT chain}

%-------------------------------------------- 
\begin{figure}%[htbp]
\begin{center}
\includegraphics[width=8.5cm]{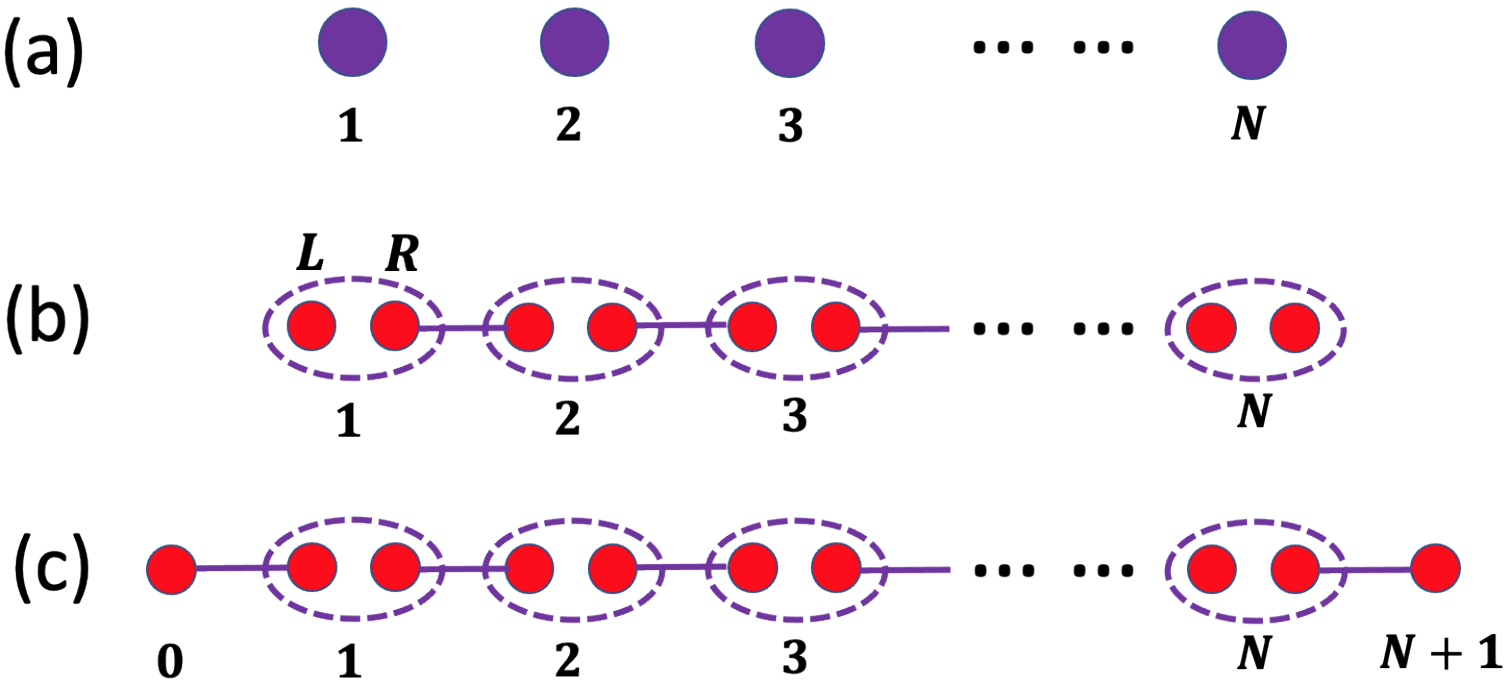}
\caption{(a) The spin-$1$ AKLT chain, (b) the ground state of AKLT chain, (c) the modified AKLT state.
In (a), the solid purple circles represent spin-$1$ sites.
In (a,b,c), the solid red circles represent spin-$1/2$ sites;
the dashed ellipses represent projection to the spin triplet sector;
and the spin-$1/2$ particles connected by a horizontal solid purple line are in a singlet state. 
} \label{fig:AKLT}
\end{center}
\end{figure}
%--------------------------------------------

We start by giving a brief review of the AKLT state. 
As shown in Fig. \ref{fig:AKLT} (a), the AKLT model is a chain of spin-$1$ particles  on a system of $N$ sites, defined by the following Hamiltonian
\bea
H_{AKLT}&=&\sum_{i=1}^{N-1}  \big[\frac{1}{2} \vec{S}_i\cdot \vec{S}_{i+1}+\frac{1}{6} (\vec{S}_i\cdot \vec{S}_{i+1})^2+\frac{1}{6}\big],
\label{eq:AKLT_ham}
\eea
in which $\vec{S}_i$ are the spin-$1$ operators on site $i$.
Alternatively, the AKLT Hamiltonian can be written as
\bea
H_{AKLT}&=& \sum_{i=1}^{N-1}P^{(2)}_{i,i+1},
\label{eq:AKLT_ham2}
\eea
in which $P^{(2)}_{i,i+1}$ is the projection to the spin-$2$ subspace of the nine-dimensional tensor product space formed by the spin-$1$ particles at sites $i$ and $i+1$. 

The ground state of $H_{AKLT}$ in Eq. (\ref{eq:AKLT_ham}) is exactly solvable, 
which can be constructed in the following way schematically shown in Fig. \ref{fig:AKLT} (b). 
First, each spin-$1$ site $i$ is fractionalized into a composition of two spin-$1/2$ sites $(i,L)$ and $(i,R)$,
where ``$L$" and ``$R$" are ``left" and ``right" for short.  
Second, the spin-1/2 particles at $(i,R)$ and $(i+1,L)$ are paired into a singlet state $\frac{1}{\sqrt{2}}(\ket{0}_{i,R}\ket{1}_{i+1,L}-\ket{1}_{i,R}\ket{0}_{i+1,L})$,
except $(1,L)$ and $(N,R)$. 
Third, since the tensor product of two spin-$1/2$ spaces contains a singlet sector and a triplet sector, a projection to the triplet sector needs to be performed on each spin-$1$ site so that the unphysical singlet sector is removed. 
We note that there is no distinction between the two fractionalized spin-$1/2$ particles on any spin-$1$ site after the projection,
since any wavefunction in the triplet sector is symmetric with respect to the exchange of the two spin-$1/2$ particles.
Hence, the symbols $L$ and $R$ are just for notational convenience, and there is  no absolute meaning for ``left" and ``right" in a spatial sense. 

Using the above described construction, the AKLT wavefuction can be written as
%\begin{widetext}
\begin{flalign}
&\ket{AKLT}= \mathcal{N} (\Pi_{i=1}^{N} P^{(1)}_i)\nn\\
&\big(\ket{\psi_{1L}}_{1L} \otimes_{i=1}^{N-1}\ket{\text{singlet}}_{(i,R),(i+1,L)}
\otimes\ket{\psi_{NR}}_{NR}\big),
\label{eq:AKLT_wf}
\end{flalign}
%\end{widetext}
in which the singlet pair $\ket{\text{singlet}}_{(i,R),(i+1,L)}$ is defined as
\begin{flalign}
\ket{\text{singlet}}_{(i,R),(i+1,L)}=\frac{1}{\sqrt{2}}(\ket{0}_{i,R}\ket{1}_{i+1,L}-\ket{1}_{i,R}\ket{0}_{i+1,L});
\end{flalign}
$\mathcal{N}$ is the normalization factor;
$P^{(1)}_i$ is the projection to the spin-$1$ subspace of the four-dimensional tensor product space formed by spin-$1/2$ particles located at $(i,L)$ and $(i,R)$;
and $\ket{\psi_{1L}}_{1L}$ and $\ket{\psi_{NR}}_{NR}$ are arbitrary spin-$1/2$ spinor wavefunctions at spin-1/2 sites $(1,L)$ and $(N,R)$, respectively. 

Notice that the total angular momentum of the two spin-$1$ particles   at sites $i$ and $i+1$ can be at most $0$ or $1$  in the wavefunction $\ket{AKLT}$.
This assertion can be easily seen by temporarily removing the projection operator $\Pi_{i=1}^N P^{(1)}_i$:
The spin-1/2 particles at sites $(i,R)$ and $(i+1,L)$ are already in a singlet state having no contribution to the total angular momentum, 
so only the spin-$1/2$ particles at  $(i,L)$ and $(i+1,R)$ can contribute, whose total angular momentum is $0$ or $1$. 
The analysis holds when $\Pi_{i=1}^N P^{(1)}_i$ is taken into account as well,
since the operator $(\vec{S}_i+\vec{S}_{i+1})^2$ commutes with the projection operator $P_j^{(1)}$ for any $j$.
Therefore, in the wavefunction $\ket{AKLT}$, there is no spin-$2$ component in the combined two-site system formed by the spin-$1$ particles at sites $i$ and $i+1$. 
Consequently, Eq. (\ref{eq:AKLT_wf}) is able to minimize the energy of Eq. (\ref{eq:AKLT_ham2}) term by term,
and as a result, it must be a ground state of the system. 
Notice that because of the arbitrariness of $\ket{\psi_{1L}}_{1L}$ and $\ket{\psi_{NR}}_{NR}$, the ground states are four-fold degenerate, corresponding to different boundary conditions. 

The ground state degeneracy can be removed by attaching two additional  spin-$1/2$ particles to the spin-$1$ chain and coupling them to spin-$1$ particles at sites $1$ and $N$ via an antiferromagnetic Heisenberg interaction as shown in Fig. \ref{fig:AKLT} (c). 
We number these two additional spin-$1/2$ sites as $0$ and $N+1$. 
The Hamiltonian now becomes 
\bea
H^\prime_{AKLT}=H_{AKLT}+J_0\vec{\sigma}_0\cdot \vec{S}_1+ J_{N+1}\vec{S}_N\cdot \vec{\sigma}_{N+1},
\label{eq:AKLT_ham_modified}
\eea
in which $\vec{\sigma}_0$ and $\vec{\sigma}_{N+1}$ are the Pauli matrices acting in the spin-$1/2$ spaces at sites $0$ and $N+1$, respectively,
and $J_0$ and $J_{N+1}$ can be any positive real number.
This time, the ground state wavefunction can be written as
\begin{flalign}
&\ket{AKLT^\prime}= \mathcal{N}^\prime (\Pi_{i=1}^{N} P^{(1)}_i)\nn\\
&\big(\ket{\text{singlet}}_{0,1L} \otimes_{i=1}^{N-1} \ket{\text{singlet}}_{(i,R),(i+1,L)}\otimes \ket{\text{singlet}}_{NR,N+1}\big),
\label{eq:AKLT_wf_prime}
\end{flalign}
in which $\mathcal{N}^\prime$ is the normalization factor. 
We will consider this modified unique AKLT state $\ket{AKLT^\prime}$ in this work. 
That $\ket{AKLT^\prime}$ minimizes the $H_{AKLT}$ term can be shown in the same way as before.
We verify that $\ket{AKLT^\prime}$ also minimizes $\vec{\sigma}_0\cdot \vec{S}_1$ and the analysis for minimizing $\vec{S}_N\cdot \vec{\sigma}_{N+1}$ is similar. 
The total angular momentum of $\frac{1}{2}\sigma_0+\vec{S}_1$ can be $1/2$ or $3/2$, according to the angular momentum addition rule. 
Since we are considering an antiferromagnetic coupling $\vec{\sigma}_0\cdot \vec{S}_1$,
the spin-$1/2$ sector has a lower energy than the spin-$3/2$ sector. 
Let's temporarily remove the projection operator $\Pi_{i=1}^{N} P^{(1)}_i$, and consider the three spin-$1/2$ particles located at $0$, $(1,L)$ and $(1,R)$.
Since the spin-$1/2$ particles at sites $0$ and $(1,L)$ already form a singlet, 
the total angular momentum of the three spin-1/2 particles comes from the spin-$1/2$ site at $(1,R)$,
which cannot have any component in the spin-$3/2$ sector. 
The analysis holds when  $\Pi_{i=1}^{N} P^{(1)}_i$ is added, since $(\frac{1}{2}\vec{\sigma}_0+\vec{S}_1)^2$ commutes with $P^{(1)}_j$ for any $j$.

It can be verified that $\ket{AKLT^\prime}$ is invariant under the following transformations ($\alpha=x,y,z$)
\bea
U_\alpha=-\sigma_0^\alpha \big(\Pi_{j=1}^{N} e^{i\pi S_j^\alpha}\big) \sigma_{N+1}^\alpha.
\label{eq:sym}
\eea
Define $\ket{\Phi}$ to be
\begin{flalign}
&\ket{\Phi}=\nn\\
&\ket{\text{singlet}}_{0,1L} \otimes_{i=1}^{N-1} \ket{\text{singlet}}_{(i,R),(i+1,L)}\otimes \ket{\text{singlet}}_{NR,N+1}.
\end{flalign}
Since a singlet pair  $\ket{\text{singlet}}_{(i,R),(i+1,L)}$ is invariant under the rotation
\bea
e^{\frac{1}{2}i(\sigma^z_{(i,R)}+\sigma^z_{(i+1,L)})\pi}=-\sigma^z_{(i,R)}\sigma^z_{(i+1,L)},
\eea
the state $\ket{\Phi}$ satisfy 
\bea
\Pi_{i=0}^{N}\big(-\sigma^z_{(i,R)}\sigma^z_{(i+1,L)}\big)\ket{\Phi}=\ket{\Phi},
\label{eq:sym_Phi}
\eea
in which $(0,R)$ and $(N+1,L)$ are identified with $0$ and $N+1$, respectively. 
The operator $\Pi_{i=0}^{N}\big(-\sigma^z_{(i,R)}\sigma^z_{(i+1,L)}\big)$ can be rewritten as
\bea
&&\Pi_{i=0}^{N}\big(-\sigma^z_{(i,R)}\sigma^z_{(i+1,L)}\big)\nn\\
&=&-\sigma^z_{0}\Pi_{j=1}^N\big(-\sigma^z_{(j,L)}\sigma^z_{(j,R)}\big)\sigma^z_{N+1}\nn\\
&=&U_\alpha.
\eea
Since $U_\alpha$ commutes with the projection operator $\Pi_{i=1}^{N} P^{(1)}_i$, Eq. (\ref{eq:sym_Phi}) implies $U_\alpha \ket{AKLT^\prime}=\ket{AKLT^\prime}$.

Notice that $U_\alpha$ mutually commute, and they also commute with the Hamiltonian in Eq. (\ref{eq:AKLT_ham_modified}).
Hence, the group generated by $\{U_\alpha\}$ is the $\mathbb{Z}_2\times \mathbb{Z}_2$ symmetry group of the system. 

%------------------------------------------------------------------------------------------------------------------------------
\subsection{The quantum teleportation protocol}
\label{subsec:teleportation}

In the standard quantum teleportation protocol \cite{Bennett1993}, Alice can teleport the state $\ket{\psi}_a$ to Bob by performing two-qubit measurements,
if they share an entangled pair of qubits $b$ and $c$, which form a Bell state $\ket{\text{singlet}}_{b,c}$.
More precisely, qubits $a$ and $b$ are on Alice's side, and qubit $c$ is on Bob's side;
the initial wavefunction before measurement is 
\bea
\ket{\text{before}}_{abc}=\ket{\psi}_a\ket{\text{singlet}}_{bc};
\eea
and Alice will measure the qubits $a$ and $b$ in the following four Bell basis states
\bea
\ket{S=0}_{ab}&=&\frac{1}{\sqrt{2}}(\ket{0}_a\ket{1}_b-\ket{1}_a\ket{0}_b)\nn\\
\ket{S=1,S^x=0}_{ab}&=&\frac{i}{\sqrt{2}}(\ket{0}_a\ket{0}_b-\ket{1}_a\ket{1}_b)\nn\\
\ket{S=1,S^y=0}_{ab}&=&\frac{1}{\sqrt{2}}(\ket{0}_a\ket{0}_b+\ket{1}_a\ket{1}_b)\nn\\
\ket{S=0,S^z=0}_{ab}&=&\frac{1}{\sqrt{2}}(\ket{0}_a\ket{1}_b+\ket{1}_a\ket{0}_b),
\label{eq:Bell_basis}
\eea
in which $\ket{S=1,S^\alpha=0}_{ab}$ ($\alpha=x,y,z$) is the eigenstate of the operator $S^\alpha$ with zero eigenvalue in the triplet sector of qubits $a$ and $b$.
The measurement is a projective measurement since the four basis states in Eq. (\ref{eq:Bell_basis}) are orthogonal to each other. 
For later convenience, we label the four states in Eq. (\ref{eq:Bell_basis}) as
\bea
\ket{s}_{ab}&=&\ket{S=0}_{ab}\nn\\
\ket{x}_{ab}&=&\ket{S=1,S^x=0}_{ab}\nn\\
\ket{y}_{ab}&=&\ket{S=1,S^y=0}_{ab}\nn\\
\ket{z}_{ab}&=&\ket{S=1,S^z=0}_{ab}.
\label{eq:Bell_basis_2}
\eea
After the measurement, the state on qubits $a$ and $b$ collapses into one of the four states $\ket{\alpha}$ ($\alpha\in\{s,x,y,z\}$) in Eq. (\ref{eq:Bell_basis}), 
and it can be verified that the state on qubit $c$ becomes $\sigma^\alpha\ket{\psi}$. 
Hence the wavefunction after Alice's measurement is 
\bea
\ket{\text{After}}=\ket{\alpha}_{ab}(\sigma^\alpha\ket{\psi})_c.  
\eea
As can be seen, the state $\psi$ on qubit $a$ has been teleported to qubit $c$, up to a byproduct operator $\sigma^\alpha$,
which is determined by the measurement result. 

The above teleportation protocol can be generalized by tilting the measurement basis.
Let's consider a modified version of quantum teleportation by measuring qubits $a$ and $b$ in a rotated basis.
The rotated basis states are obtained by acting $e^{i S^\gamma_{ab} \theta}$ ($\gamma\in\{x,y,z\}$) on the four states in Eq. (\ref{eq:Bell_basis}),
in which $S^\gamma_{ab}=\frac{1}{2}(\sigma^\gamma_a+\sigma^\gamma_b)$.  
For example, if $\gamma=z$, the four rotated basis states become
\bea
\ket{s,(z,\theta)}&=&\ket{S=0}\nn\\
\ket{x,(z,\theta)}&=&\cos(\frac{\theta}{2})\ket{S=1,S^x=0}\nn\\
&&+\sin(\frac{\theta}{2})\ket{S=1,S^y=0}\nn\\
\ket{y,(z,\theta)}&=&-\sin(\frac{\theta}{2})\ket{S=1,S^x=0}\nn\\
&&+\cos(\frac{\theta}{2})\ket{S=1,S^y=0}\nn\\
\ket{z,(z,\theta)}&=&\ket{S=1,S^z=0},
\label{eq:Bell_modified}
\eea
in which the subscript $ab$ has been omitted, and the overall phase factor in $\ket{z,(z,\theta)}=e^{iS^z\theta}\ket{S=1,S^z=0}$ has been dropped.
The basis states after $x$- and $y$-rotations can be obtained from Eq. (\ref{eq:Bell_modified}) by permuting $x,y,z$.
After measuring qubits $a$ and $b$ in the rotated basis, 
while the wavefunction on qubits $a$ and $b$ collapses to $\ket{\alpha,(z,\theta)}_{ab}$ where $\alpha\in\{s,x,y,z\}$, 
it can be verified that  the state of qubit $c$ becomes $(\sigma^\alpha e^{iS^\gamma\theta}\ket{\psi})_c$ when $\alpha\in\{x,y\}$
and becomes $(\sigma^z \ket{\psi})_c$ when $\alpha=z$. 
To summarize, the  wavefunction of the three qubits after measurement is 
\begin{flalign}
\ket{\text{After}}_{abc}=
\begin{cases}
\ket{\alpha,(z,\theta)}_{ab}(\sigma^\alpha e^{i\frac{1}{2}\sigma^\gamma\theta}\ket{\psi})_c,& \alpha\in\{x,y\},\\
\ket{z}_{ab}(\sigma^z \ket{\psi})_c,& \alpha=z.
\end{cases}
\label{eq:state_after}
\end{flalign}
As can be seen from Eq. (\ref{eq:state_after}), by measuring the four states in the rotated basis defined in Eq. (\ref{eq:Bell_modified}) on qubits $a$ and $b$,
the state on qubit $a$ is teleported to qubit $c$ up to a rotation around $z$-axis and a byproduct operator. 

%------------------------------------------------------------------------------------------------------------------------------
\subsection{MBQC in AKLT state}
\label{sec:MBQC_AKLT}

We briefly review how to implement MBQC using the modified AKLT state $\ket{AKLT^\prime}$ defined in Eq. (\ref{eq:AKLT_wf_prime}).
The purpose is to show that any logical SU(2) rotation can be implemented in MBQC. 
We will show in an inductive manner the method for implementing a rotation around $x$-, $y$- or $z$-axis in MBQC. 
Using the Euler decomposition of an arbitrary rotation, the statement follows directly. 

First, by measuring site $0$ in the basis of $\sigma^\alpha$-eigenstates ($\alpha\in\{x,y,z\}$), the fractionalized spin-1/2 particle at position $(1,L)$ can be initialized to $\sigma^\alpha$ eigenstates up to a byproduct operator. 
We consider a $\sigma^x$ measurement as an example, and the analysis for $\sigma^y$ and $\sigma^z$ measurements are similar. 
Let's drop the projection operator $\Pi_{i=1}^N P^{(1)}_i$ temporarily as before. 
Let $s_0\in\{0,1\}$ be the measurement outcome of the $\sigma^x$-measurement on site $0$,
so that the wavefunction on site $0$ collapses to $\ket{+}$ ($\ket{-}$) for $s_0=0$ ($s_0=1$).
Since the spin-$1/2$ particles at positions $0$ and $(1,L)$ form a singlet state $\frac{1}{\sqrt{2}}(\ket{+}_0\ket{-}_{1L}-\ket{-}_0\ket{+}_{1L})$,
%the state at position $(1,L)$ is fixed to be $\ket{-}$ ($\ket{+}$) for $s_0=0$ ($s_0=1$) after the measurement.  
the state at $(1,L)$ is initialized to the $\ket{+}$ state up to an overall sign when $s_0=0$;
whereas when $s_0=1$, it is initialized to $\ket{-}=\sigma^z\ket{+}$,
hence  still becomes the $\ket{+}$ state up to a byproduct operator $\sigma^z$ (note: choosing the byproduct operator to be $\sigma^y$ works equally well).
Therefore, we see that the state at $(1,L)$ is $(\sigma^z)^{s_0}\ket{+}$ after measuring site $0$,
thereby is initialized to $\ket{+}$ state up to the byproduct operator $(\sigma^z)^{s_0}$. 
Notice that adding $\Pi_{i=1}^N P^{(1)}_i$ does not change the conclusion, since the projection operators $\ket{+}_0\bra{+}$ and $\ket{-}_0\bra{-}$ on site $0$ commute with $\Pi_{i=1}^N P^{(1)}_i$. 

Next, the $N$ spin-$1$ sites are sequentially measured, with the following measurement outcomes $s_j$ and  the corresponding measurement basis at site $j$,  
\bea
s_j=x&:&\ket{x,(\gamma_j,\theta_j)}_j=e^{iS_j^{\gamma_j}\theta_j}\ket{S^x=0}_j,\nn\\
s_j=y&:&\ket{y,(\gamma_j,\theta_j)}_j=e^{iS_j^{\gamma_j}\theta_j}\ket{S^y=0}_j,\nn\\
s_j=z&:&\ket{z,(\gamma_j,\theta_j)}_j=e^{iS_j^{\gamma_j}\theta_j}\ket{S^z=0}_j,
\label{eq:rotated_basis}
\eea 
where $\gamma_j\in\{x,y,z\}$ and $\theta_j\in[0,2\pi)$. 
For later convenience, we abbreviate the three states in Eq. (\ref{eq:rotated_basis}) simply as $\ket{\alpha}_j$ ($\alpha\in\{x,y,z\}$) when $\theta_j=0$.
Notice that when $s_j=\gamma_j$, $e^{iS_j^{\gamma_j}\theta_j}\ket{S^{\gamma_j}=0}_j$ is just $\ket{S^{\gamma_j}=0}_j$ up to an overall phase.
In practice, the choices of the direction $\gamma_j$ and the angle $\theta_j$ depend on specific quantum algorithms and measurement outcomes at sites prior to site $j$, which we explain below. 

Suppose site $0$ is measured in the $\sigma^x$-basis;
spin-$1$ sites $1$ to $k-1$ have been measured; 
and the algorithm demands the implementation of a rotation of angle $\varphi$ around $\gamma$-axis ($\gamma\in\{x,y,z\}$) in the next step.  
Let $\ket{\psi_{k-1}}_{(k,L)}$ be the state of the spin-$1/2$ particle at position $(k,L)$ at the current stage. 
Now we consider spin-$1/2$ sites $(k,L)$, $(k,R)$ and $(k+1,L)$. 
Because of the construction of the AKLT state, the spin-$1/2$ particles at $(k,R)$ and $(k+1,L)$ are in a singlet state. 
Therefore, the measurement of spin-$1$ site $k$ (i.e., the combined spin-$1/2$ sites $(k,L)$ and $(k,R)$) is exactly the quantum teleportation protocol described in Sec. \ref{subsec:teleportation},
except that now the $\ket{S=0}$ basis state in Eq. (\ref{eq:Bell_modified}) does not show up in the measurement since the unphysical singlet component has been projected out in the construction of the AKLT state. 
In the present case, the spin-$1/2$ sites $(k,L)$ and $(k,R)$ belong to Alice, and $(k+1,L)$ belongs to Bob.
Suppose the measurement angle is chosen as $\theta_k$ (the relation between $\theta_k$ and $\varphi$ will be given below in Eq. (\ref{eq:thetak_varphi})), then by
directly borrowing the discussion from Sec. \ref{subsec:teleportation},
we see that after the measurement, 
$(k,L)$ and $(k,R)$ collapse to one of the three basis states $\{e^{i S_k^\gamma \theta_k}\ket{S^{s_k}=0}_k~|~s_k\in\{x,y,z\}\}$, where $s_k$ is the measurement outcome;
whereas the wavefunction $\ket{\psi_k}_{k+1,L}$ at $(k+1,L)$ becomes $\big(\sigma^{s_k} e^{i\frac{1}{2}\sigma^\gamma\theta_k}\ket{\psi_{k-1}}\big)_{k+1,L}$
when $s_k\neq \gamma$, and becomes $\big(\sigma^{\gamma}\ket{\psi_{k-1}}\big)_{k+1,L}$ when $s_k=\gamma$.
Namely, in an inductive language, the information-carrying qubit has shifted from site $(k,L)$ to $(k+1,L)$, and $\ket{\psi_k}$ is related to $\ket{\psi_{k-1}}$ via
\bea
\ket{\psi_k}=
\begin{cases}
\sigma^{s_k} e^{i\frac{1}{2}\sigma^\gamma\theta_k}\ket{\psi_{k-1}},& s_k\neq \gamma\nn\\
\sigma^{\gamma}\ket{\psi_{k-1}},&s_k=\gamma.
\end{cases}
\eea

However, notice that the measurement angle $\theta_k$ may be different from the algorithm angle $\varphi$ because of the byproduct operators $(\sigma^z)^{s_0}$ ($s_0\in\{0,1\}$) and $\sigma^{s_j}$ ($s_j\in\{x,y,z\}$, $1\leq j\leq k-1$) generated by measuring the sites before site $k$. 
Because of the anti-commutation relations between different Pauli matrices,
the measurement angle needs to adjust a sign when $s_j\neq \gamma$, for $1\leq j\leq k-1$.  
Site $0$ is special and requires a separate treatment. 
Since the byproduct operator is $(\sigma^z)^{s_0}$,
a sign needs to be inserted in the measurement angle if $\gamma\neq z$ and $s_0=1$. 
Therefore,  in order to implement a $\varphi$-rotation around $\gamma$-direction, we need to choose the measurement angle $\theta_k$ as
\bea
\theta_k=\varphi (-)^{s_0(1-\delta_{\gamma,z})}\Pi_{j=1}^{k-1}(-)^{1-\delta_{\gamma,s_j}},
\label{eq:thetak_varphi}
\eea
which reflects the adaptive nature of MBQC. 

If $s_k\neq \gamma$, a rotation around $\gamma$-direction has been successfully implemented. 
If $s_k= \gamma$, no rotation is implemented at the step of site $k$. 
One needs to move onto site $k+1$ and repeat the same measurement protocol as was done at site $k$.
This time, the measurement angle needs to be chosen as 
$\theta_{k+1}=\varphi (-)^{s_0(1-\delta_{\gamma,z})}\Pi_{j=1}^{k}(-)^{1-\delta_{\gamma,s_k}}$. 
Since the probability of failing to implement the rotation decays exponentially as the number of sites being measured increases,
the success rate of implementing a $\varphi$-rotation around $\gamma$-axis can be made arbitrarily close to unity.  

When sites $0$ to $N$ have all been measured, the entanglements in the AKLT state have been destroyed.
The information-carrying qubit shifts  to the spin-$1/2$ site at position $N+1$,
which is related to the initialized logical state $\ket{+}$ by a sequence of SU(2) rotations up to a product of byproduct operators. 

%%%%%%%%%%%%%%%%%%%%%%%%%%%%%%%%%%%%%%%%%%%%%%%%%%%%%%
\section{SPT-MBQC in Haldane phase by summing over all measurement paths}
\label{sec:sum_paths}

\begin{figure}%[htbp]
\begin{center}
\includegraphics[width=8.7cm]{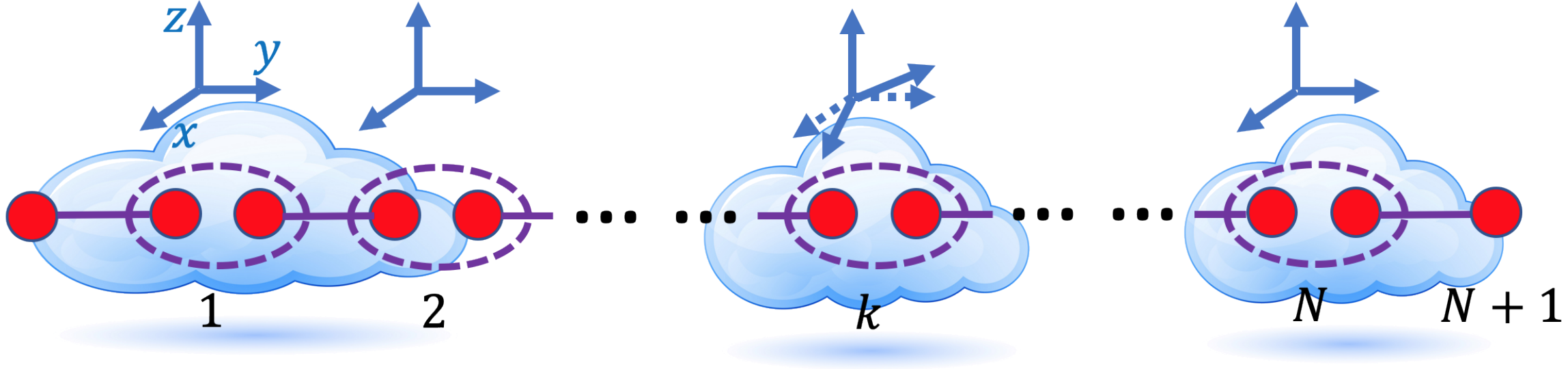}
\caption{Illustration of the measurement setting for realizing a single logical rotation in SPT-MBQC located at site $k$ as discussed in Sec. \ref{sec:SPT_MBQC_sum_path}.
The basis states for spin-$1$ sites in the measurements are chosen to be $\{\ket{\alpha}~|~ \alpha=x,y,z\}$ except site $k$, where the basis states are rotated by an angle $\theta_k$ around the $z$-axis. 
The clouds are used to indicate the fact that for general systems in the SPT phase beyond the AKLT point, the fractionalized particles get  smeared, extending to a range of sites in the chain. 
} \label{fig:single_rot_SPT} 
\end{center}
\end{figure}

In this section, by summing over all measurement paths \cite{Raussendorf2017}, we show that MBQC can be performed in the spin-$1$ Haldane phase, not just at the AKLT point. 
We will consider the realization of a single logical rotation located at a fixed site in the chain as shown in Fig. \ref{fig:single_rot_SPT}.
The generalization to arbitrary number of rotations and any qudit system will be discussed in Sec. \ref{sec:algebraic_MBQC} using an algebraic approach to SPT-MBQC. 
The central result of this section is summarized in Proposition \ref{proposition:single_rot} below,
which will be derived in Sec. \ref{sec:SPT_MBQC_sum_path}.

As in Sec. \ref{sec:review}, we number the spin-$1$ sites as $\{1,2,...,N\}$,
and the spin-$1/2$ sites as $0$ and $N+1$. 

\begin{Prop}\label{proposition:single_rot}
Let $\ket{G}$ be a ground state in the Haldane phase of a spin-$1$ chain with two additional spin-$1/2$ particles attached to the left and right boundaries. 
Suppose site $0$ is measured in the $\sigma^x$-basis,
and all spin-$1$ sites are measured in the unrotated basis, except site $k$ ($1\leq k\leq N$) which is measured in the basis rotated around $z$-axis characterized by angle $\varphi$. 
Then the expectation values of measuring $\sigma^\alpha$ at site $N+1$ (denoted as $\langle\langle\sigma^\alpha_{N+1}\rangle\rangle$) obtained by averaging over rounds of MBQC simulations are given by
\bea
\langle\langle\sigma^x_{N+1}\rangle\rangle&=&\bra{G}\cos(S_k^z \varphi)\ket{G}\nn\\
\langle\langle\sigma^y_{N+1}\rangle\rangle&=&\bra{G} \sin(S_k^z\varphi) \big(\Pi_{j=k}^Ne^{i\pi S_j^z}\big)\sigma_{N+1}^z\ket{G}\nn\\
\langle\langle\sigma^z_{N+1}\rangle\rangle&=&0.
\label{eq:z_rotate_1}
\eea
\end{Prop} \medskip

In Eq. (\ref{eq:z_rotate_1}), $\langle\langle\sigma^\alpha_{N+1}\rangle\rangle$ is used to denote the average of the usual expectation value over all measurement paths (for details, see Eq. (\ref{eq:sum_path_expr})). 

We note that Eq. (\ref{eq:z_rotate_1}) acquires a simpler form in the $\varphi\ll 1$ limit. 
Neglecting second and higher order terms of $\varphi$, Eq. (\ref{eq:z_rotate_1}) reduces to 
\bea
\langle\langle\sigma^x_{N+1}\rangle\rangle&=&1\nn\\
\langle\langle\sigma^y_{N+1}\rangle\rangle&=&\nu_z\varphi\nn\\
\langle\langle\sigma^z_{N+1}\rangle\rangle&=&0,
\eea
where $\nu_z$ is the bulk-to-end string order parameter defined as
\bea
\nu_z=\bra{G} S_k^z \big(\Pi_{j=k}^Ne^{i\pi S_j^z}\big)\sigma_{N+1}^z\ket{G}. 
\label{eq:nu_z}
\eea
When $\varphi=0$, we have $\langle\langle\sigma^x_{N+1}\rangle\rangle=1$ and $\langle\langle\sigma^y_{N+1}\rangle\rangle=\langle\langle\sigma^z_{N+1}\rangle\rangle=0$,
corresponding to a logical $X$-eigenstate with eigenvalue $1$,
which is consistent with our expectation from the initialization procedure. 
On the other hand, when $\varphi$ is nonzero and in the small $\varphi$ limit, $\langle\langle\sigma^\alpha_{N+1}\rangle\rangle$'s are approximately given by a rotation around $z$-axis, with a renormalized rotation angle  $\nu_z\varphi$, different from $\varphi$. 

Proposition \ref{proposition:single_rot} demonstrates the ability of doing MBQC in the SPT phase for approximately implementing a $z$-rotation. 
Therefore, it implies that SPT-MBQC can be realized  without referring to projective representations in the bulk,
which do not exist for integer spins. 

%-------------------------------------------- 
%--------------------------------------------

%------------------------------------------------------------------------------------------------------------------------------------   
\subsection{The Haldane phase}

We start by giving a brief review of 1D symmetry protected topological (SPT) phases. 
Two systems are said to be in the same SPT phase if they satisfy:
(1) The ground states of the two systems are both gapped without any symmetry breaking;
(2) the two ground states can be adiabatically connected to each other via symmetry-preserving local unitary transformations without gap closing. 

%-------------------------------------------- 
%\begin{figure}%[htbp]
%\begin{center}
%\includegraphics[width=8cm]{string_order.pdf}
%\caption{String order parameter defined in Eq. (\ref{eq:string_order}) for $i=1$, $j=3$.
%} \label{fig:string_order}
%\end{center}
%\end{figure}
%--------------------------------------------

As an example, instead of $H^\prime_{AKLT}$ in Eq. (\ref{eq:AKLT_ham_modified}), we consider the following more general Hamiltonian on a system of $N$ spin-$1$ sites and two additional spin-$1/2$ sites at $0$ and $N+1$, 
\bea
H&=& \sum_{i=1}^{N-1} \big[ \cos(\theta) \vec{S}_i\cdot \vec{S}_{i+1}+\sin(\theta) (\vec{S}_i\cdot \vec{S}_{i+1})^2\big]\nn\\
&&+ D_x \sum_{i=1}^{N-1} (S_i^x)^2+D_z \sum_{i=1}^{N-1} (S_i^z)^2 \nn\\
&&+J_0\vec{\sigma}_0\cdot \vec{S}_1+ J_{N+1}\vec{S}_N\cdot \vec{\sigma}_{N+1}.
\label{eq:haldane_general}
\eea
which reduces to $H^\prime_{AKLT}$ when $D_x=D_z=0$, and $\theta=\arctan(1/3)$. 
Unlike $H^\prime_{AKLT}$ which is SU(2) invariant, it can be easily verified that $H$ in Eq. (\ref{eq:haldane_general}) has a global $\mathbb{Z}_2\times\mathbb{Z}_2$ symmetry generated by $R(\hat{x},\pi)$ and $R(\hat{z},\pi)$, where $R(\hat{n},\beta)=e^{\frac{1}{2} (\vec{\sigma}_0+\vec{\sigma}_{N+1})\cdot \hat{n} \beta}\Pi_{i=1}^N e^{i\vec{S}_i\cdot\hat{n}\beta}$ denotes a global spin rotation around $\hat{n}$-direction by an angle $\beta$.
For certain ranges of $\theta$, $D_x$ and $D_z$, $H$ is in the same SPT phase as $H^\prime_{AKLT}$ protected by the above mentioned $\mathbb{Z}_2\times\mathbb{Z}_2$ symmetry. 
For example, if $D_x=D_z=0$,  
all Hamiltonians for $\theta\in(-\pi/4,\pi/4)$ are in the same SPT phase, known as the Haldane phase.

A useful quantity to characterize the spin-$1$ Haldane phase is the string order parameter $\mathcal{O}^\alpha_{ij}$ ($\alpha=x,y,z$) defined as ($i<j$)
\bea
\mathcal{O}^\alpha_{ij}=\langle S_i^\alpha (\Pi_{m=i+1}^{j-1} e^{i\pi S_m^\alpha}) S_j^\alpha \rangle,~|i-j|\rightarrow \infty,
\label{eq:string_order}
\eea
which is non-vanishing (vanishing) in (out of) the Haldane phase. 
The sites $i,j$ are both in the bulk, hence Eq. (\ref{eq:string_order}) can be viewed as bulk-to-bulk string order parameter.
In MBQC, 
as can be seen from the discussions following Eq. (\ref{eq:nu_z}),  
what turns out to be useful is the bulk-to-end string order parameter defined as
\bea
\mathcal{O}^\alpha_{i,N+1}=\langle S_i^\alpha (\Pi_{m=i+1}^{N} e^{i\pi S_m^\alpha}) \sigma^\alpha_{N+1} \rangle. 
\label{eq:string_bulk_end}
\eea

%-------------------------------------------- 
\begin{figure}%[htbp]
\begin{center}
\includegraphics[width=8.3cm]{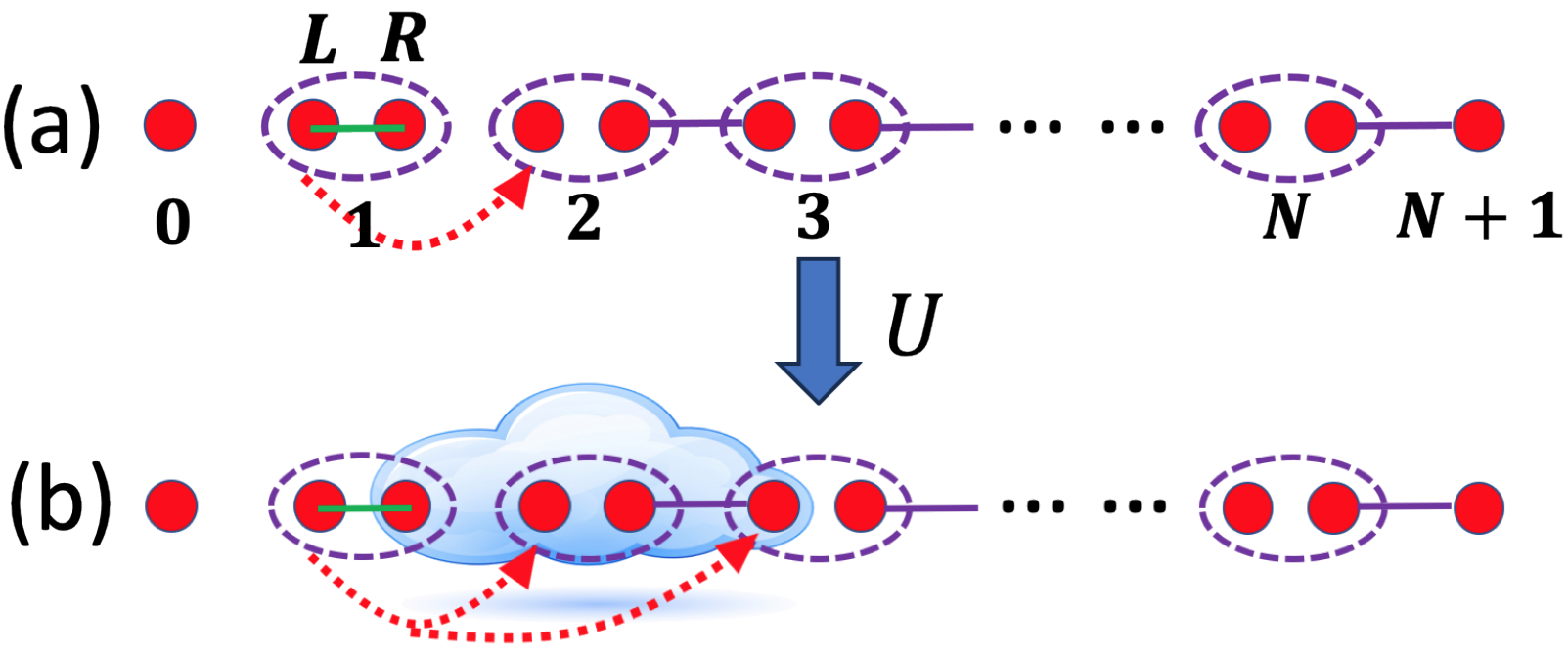}
\caption{(a) Transfer of information-carrying qubit in MBQC in the AKLT chain, (b) a many-body state in the Haldane phase which is related to the AKLT state by a local unitary transformation $U$ preserving the $\mathbb{Z}_2\times \mathbb{Z}_2$ symmetry and the gap. 
In (a,b), the horizontal green line represents one of the three states $\ket{S=1,S^\alpha=0}$ ($\alpha\in\{x,y,z\}$) in the triplet sector, which is the state the spin-$1$ particle at site $1$ collapses to after measuring the first spin-$1$ site.
In (a), the dashed arrowed line denotes the transfer of information-carrying qubit from $(1,L)$ to $(2,L)$.
In (b), the cloud is used to illustrate the fact that the fractionalized spin-$1/2$ particle gets smeared after the transformation $U$ is applied to the AKLT state;
and the multiple dashed arrowed lines are used to indicate the situation where the transferred information is no longer located at a single fictitious spin-$1/2$ site, but extends to a finite range of sites. 
} \label{fig:AKLT_smeared} 
\end{center}
\end{figure}
%--------------------------------------------

It is intuitively not difficult to understand why the method described in Sec. \ref{sec:MBQC_AKLT} for realizing MBQC does not work in the Haldane phase, away from the AKLT point.
According to the notion of SPT phase, the ground state $\ket{G}$ of a general state in the Haldane phase can be obtained from the AKLT state by a symmetry-preserving local unitary transformation $U$ without gap closing, i.e., 
\bea
\ket{G}=U\ket{AKLT^\prime}.
\eea
As discussed in Sec. \ref{sec:MBQC_AKLT} and shown in Fig. \ref{fig:AKLT_smeared} (a), 
the information-carrying qubit is transferred from qubit $(1,L)$ to $(2,L)$ after measuring the spin-$1$ site $1$ in the AKLT chain.
In the state $\ket{G}$, the fractionalized spin-$1/2$ particles get smeared because of the local unitary transformation $U$,
meaning for example, $(2,L)$ is no longer located on the spin-$1$ site $1$, but extends to a finite range of sites surrounding spin-$1$ site $1$, which is schematically plotted as a cloud in Fig. \ref{fig:AKLT_smeared} (b). 
This means that one loses the control over the information-carrying qubit for general states in the SPT phase,
and a new scheme is needed to implement MBQC. 

%------------------------------------------------------------------------------------------------------------------------------------   
\subsection{SPT-MBQC by summing over all measurement paths}
\label{sec:SPT_MBQC_sum_path}
 
The above mentioned difficulty of performing MBQC in the SPT phase can be overcome by summing over all measurement paths, as discussed in Ref. \onlinecite{Raussendorf2017}.
%We will call the scheme of MBQC in the SPT phase as SPT-MBQC for short. 

In one full round of measurements in MBQC, the spin-$1/2$ particles at sites $0$ and $N+1$, as well as the spin-$1$ particles at sites $1$ to $N$ are measured in basis states determined by both the algorithm and the measurement results on other sites. 
The measurement outcomes  form an $(N+2)$-dimensional vector $(s_0,s_1,...,s_N,s_{N+1})$ where $s_0,s_{N+1}\in\{0,1\}$ and $s_i\in\{x,y,z\}$, $1\leq i\leq N$.
We call the vector $(s_0,s_1,...,s_N,s_{N+1})$ a measurement path. 
In Sec. \ref{sec:MBQC_AKLT}, MBQC in the AKLT chain has been interpreted as an evolution of the logical qubit, transferring from the fractionalized site $(1,L)$ to the last site at $N+1$. 
Here we abandon this picture, and instead revisit the procedure from an operational point of view.
In practice, suppose we want to obtain the expectation value of any hermitian operator $O_{N+1}$ at site $N+1$, which is a linear combination $\sum_{\alpha=0}^3\lambda_\alpha \sigma^\alpha_{N+1}$ of $\sigma^0_{N+1}=I_{N+1}$ and $\sigma_{N+1}^\alpha$ ($\alpha=x,y,z$), where $I_{N+1}$ is the $2\times 2$ identity operator at site $N+1$,
and $\lambda_\alpha$'s are real numbers. 
In practice, multiple rounds  of measurement need to be performed.
Let  $(s_0(i),s_1(i),...,s_N(i),s_{N+1}(i))$ be the vector of measurement outcomes at round $i$ where $1\leq i\leq M$,
and $\mu(i)$ be the result obtained from the last measurement at site $N+1$ (i.e., $\mu(i)$ is one of the two eigenvalues of the operator $O_{N+1}$).
Then when $M$ is large enough, the expectation value of $O_{N+1}$ can be obtained from 
\bea
\langle O_{N+1} \rangle = \frac{1}{M}\sum_{i=1}^M \mu(i). 
\label{eq:expectation_practical}
\eea

In MBQC performed on AKLT state, in principle only one measurement path needs to be kept (though in practice it is unfavorable to do this for the cost of time).
Namely, one can do post-selections such that only those measurement rounds $i$ which satisfy $(s_0(i),s_1(i),...,s_N(i),s_{N+1}(i))=(s_0,s_1,...,s_N,s_{N+1})$ are retained and all other measurement rounds are discarded. 
Then Eq. (\ref{eq:expectation_practical}) in this scenario reduces to
\bea
\langle O_{N+1} \rangle =\frac{1}{M_1} \sum_{\vec{s}(i)=\vec{s}} \mu(i),
\label{eq:expectation_practical_reduced}
\eea
in which $\vec{s}=(s_0,...,s_{N+1})$,  $\vec{s}(i)=(s_0(i),...,s_{N+1}(i))$, and 
$M_1=|\{i|\vec{s}(i)=\vec{s}\}|$, where $|A|$ denotes the number of elements in the finite set $A$. 

On the other hand, in SPT-MBQC, the reduced protocol of keeping only a single measurement path in Eq. (\ref{eq:expectation_practical_reduced}) is no longer applicable,
which is essentially because the transfer of information-carrying qubit along the chain loses its sense. 
For a general many-body state in the SPT phase, Eq. (\ref{eq:expectation_practical}) must be applied, 
which corresponds to summing over all measurement paths. 
Without loss of generality, we consider the special cases $O_{N+1} =\sigma^\alpha$ ($\alpha=0,x,y,z$),
where the observable $O_{N+1}$ is a Pauli operator.
The result of a general case can be obtained from a linear combination of these special cases. 

The measurement protocol is as follows. 
We fix the rotation site to be $k$ where $1\leq k\leq N$,
and consider a rotation around $\gamma$-axis $\gamma\in\{x,y,z\}$. 
Site $0$ is measured in the $\sigma^x$-eigenbasis with a byproduct operator $(\sigma^z)^{s_0}$;
sites $\{j~|~j\neq k, 1\leq j\leq N\}$ are measured in the basis $\{\ket{\alpha}=\ket{\alpha,\theta_j=0}\}_{\alpha=x,y,z}$ defined by Eq. (\ref{eq:rotated_basis});
site $k$ is measured in the rotated basis $\{\ket{\alpha,(\gamma,\theta_k)}\}_{\alpha=x,y,z}$ defined by Eq. (\ref{eq:rotated_basis}),
where  $\theta_k$ is given in Eq. (\ref{eq:thetak_varphi}). 
Define  $\ket{0,x}=\ket{+}$ and $\ket{1,x}=\ket{-}$ to be the $\sigma^x$-eigenstates. 
After measuring sites $0$ to $N$, the initial state $\ket{G}$ collapses to 
\bea
\ket{s_0,x}_0 \ket{s_1}_1...\ket{s_k,(\gamma,\theta_k)}_{k}...\ket{s_{N}}_{N} \otimes \ket{\psi}_{N+1},
\eea
in which 
\bea
\ket{\psi(\vec{s})}_{N+1}=\frac{\bra{s_0,x}_0 \bra{s_1}_1...\bra{s_k,(\gamma,\theta_k)}_{k}...\bra{s_{N}}_{N} \ket{G}}{|\bra{s_0,x}_0 \bra{s_1}_1...\bra{s_k,(\gamma,\theta_k)}_{k}...\bra{s_{N}}_{N} \ket{G}|},
\eea
where $\ket{\psi}_{N+1}$ is a function of the measurement path $\vec{s}$ highlighted by $\ket{\psi(\vec{s})}_{N+1}$. 
Suppose we want to measure $\sigma^\alpha_{N+1}$. 
For a given measurement path, the expectation value $\langle\sigma^\alpha_{N+1}\rangle$ is given by  
\begin{flalign}
&\langle\sigma^\alpha_{N+1}\rangle(\vec{s})=\nn\\
&\bra{\psi(\vec{s})} (-)^{s_0(1-\delta_{\alpha,z})}\big[\Pi_{j=1}^N(-)^{1-\delta_{\alpha,s_j}}\big]\sigma^\alpha \ket{\psi(\vec{s})}_{N+1},
\end{flalign}
in which $(-)^{s_0(1-\delta_{\alpha,z})}\big[\Pi_{j=1}^N(-)^{1-\delta_{\alpha,s_j}}\big]$ originates from the adaptive nature of MBQC, i.e., the byproduct operators $(\sigma^z)^{s_0}$ produced by measuring site $0$ and $\sigma^{s_j}$'s produced by measuring sites $j$ ($1\leq j\leq N$) can change the sign of $\sigma^\alpha_{N+1}$;
and $\langle\sigma^\alpha_{N+1}\rangle(\vec{s})$ is used to denote the fact that in general the expectation value depends on $\vec{s}$.
Here we emphasize that although the protocol in Sec. \ref{sec:MBQC_AKLT} no longer applies, the same byproduct operators are inserted to the readout qubit at site $N+1$ even in SPT-MBQC, which can be viewed as imposed by hand in the SPT-MBQC procedure.  

If all measurement paths are summed over, the expectation value $\langle\langle\sigma^\alpha_{N+1}\rangle\rangle$ is given by 
\begin{flalign}
&\langle\langle\sigma^\alpha_{N+1}\rangle\rangle=\sum_{\vec{s}}P(\vec{s})\langle\sigma^\alpha_{N+1}\rangle(\vec{s}),
\label{eq:sum_path_expr}
\end{flalign}
in which the probability $P(\vec{s})$ is 
\bea
P(\vec{s})=|\bra{s_0,x}_0 \bra{s_1}_1...\bra{s_k,(z,\theta_k)}_{k}...\bra{s_{N}}_{N} \ket{G}|^2,
\eea
and the double bracket $\langle\langle...\rangle\rangle$ is used to 
indicate averaging of the usual expectation value $\langle ... \rangle$ over all measurement paths.
%the situation where not only the usual expectation value over states but also the average over all measurement paths are taken. 

Next, we simplify the expression of $\langle\langle\sigma^\alpha_{N+1}\rangle\rangle$ in Eq. (\ref{eq:sum_path_expr}). 
We have
\begin{flalign}
&\langle\langle\sigma^\alpha_{N+1}\rangle\rangle=\sum_{\vec{s}}\bra{G}(-)^{s_0(1-\delta_{\alpha,z})}\ket{s_0,x}_0\bra{s_0,x}\nn\\
&\cdot\Pi_{j\neq k,1\leq j\leq N}(-)^{1-\delta_{\alpha,s_j}}\ket{s_j}_j\bra{s_j}\nn\\
&\cdot(-)^{1-\delta_{\alpha,s_k}}\ket{s_k,(\gamma,\theta_k)}_k\bra{s_k,(\gamma,\theta_k)}\cdot \sigma^\alpha_{N+1}\ket{G}.
\end{flalign}
Using
\bea
\ket{s_k,(\gamma,\theta_k)}\bra{s_k,(\gamma,\theta_k)}=e^{i\frac{1}{2}iS_k^\gamma\theta_k} \ket{s_k}\bra{s_k} e^{-i\frac{1}{2}iS_k^\gamma\theta_k},
\eea
\bea
\theta_k=\varphi \Pi_{j=1}^{k-1}(-)^{1-\delta_{\gamma,s_j}},
\eea
and 
\bea
e^{i\pi S_j^\gamma}\ket{s_j}_j\bra{s_j}=(-)^{1-\delta_{\gamma,s_j}}\ket{s_j}_j\bra{s_j},
\eea
$\langle\langle\sigma^\alpha_{N+1}\rangle\rangle$ can be written as 
\begin{flalign}
&\langle\langle\sigma^\alpha_{N+1}\rangle\rangle\nn\\
&=\sum_{\vec{s}}\bra{G}
(-)^{s_0(1-\delta_{\alpha,z})} \big( \Pi_{j=1}^N(-)^{1-\delta_{\alpha,s_j}}\big)
\ket{s_0,x}_0\bra{s_0,x}\nn\\
&e^{i\frac{1}{2} S_k^\gamma \varphi \Pi_{m=1}^{k-1}e^{i\pi S_m^\gamma}} \Pi_{j=1}^N\ket{s_j}_j\bra{s_j} 
e^{-i\frac{1}{2} S_k^\gamma \varphi \Pi_{m=1}^{k-1}e^{i\pi S_m^\gamma}}
\sigma^\alpha_{N+1}\ket{G}\nn\\
&=\bra{G} \big(\sum_{s_0}  (-)^{s_0(1-\delta_{\alpha,z})}\ket{s_0,x}_0\bra{s_0,x}\big)\cdot e^{i\frac{1}{2} S_k^\gamma \varphi \Pi_{m=1}^{k-1}e^{i\pi S_m^\gamma}} \nn\\
&\big(\Pi_{j=1}^N \sum_{s_j}(-)^{1-\delta_{\alpha,s_j}}\ket{s_j}_j\bra{s_j}\big)  e^{-i\frac{1}{2} S_k^\gamma \varphi \Pi_{m=1}^{k-1}e^{i\pi S_m^\gamma}}\sigma_{N+1}^\alpha\ket{G}.
\end{flalign}
Then using
\begin{flalign}
&\sum_{s_0=0,1}(-)^{s_0(1-\delta_{\alpha,z})}\ket{s_0,x}\bra{s_0,x}
=\begin{cases}
I, &\alpha=z\\
\sigma^x,&\alpha=x,y
\end{cases},\nn\\
&\sum_{s_j=x,y,z}(-)^{1-\delta_{\alpha,s_j}}\ket{s_j}\bra{s_j}=e^{i\pi S_j^\alpha},
\end{flalign}
we obtain
\bea
\langle\langle\sigma^\alpha_{N+1}\rangle\rangle&=&\bra{G}
\big[\delta_{\alpha z}+\sigma_0^x (\delta_{\alpha x}+\delta_{\alpha y})\big]e^{i\frac{1}{2} S_k^\gamma \varphi \Pi_{m=1}^{k-1}e^{i\pi S_m^\gamma}} \nn\\
&&\big(\Pi_{j=1}^Ne^{i\pi S_j^\alpha}\big)e^{-i\frac{1}{2} S_k^\gamma \varphi \Pi_{m=1}^{k-1}e^{i\pi S_m^\gamma}}\cdot\sigma^\alpha_{N+1}\ket{G},\nn\\
\label{eq:double_expectation_a}
\eea
in which
\begin{flalign}
&e^{i\frac{1}{2} S_k^\gamma \varphi \Pi_{m=1}^{k-1}e^{i\pi S_m^\gamma}} e^{i\pi S_k^\alpha}e^{-i\frac{1}{2} S_k^\gamma \varphi \Pi_{m=1}^{k-1}e^{i\pi S_m^\gamma}}\nn\\
%&=e^{\frac{1}{2}iS_k^z\theta_k} (e^{i\pi S_k^\alpha}e^{-\frac{1}{2}iS_k^z\theta_k}e^{-i\pi S_k^\alpha})e^{i\pi S_k^\alpha}\nn\\
&=\begin{cases}
e^{i\pi S_k^\alpha },&\alpha=\gamma\\
e^{i S_k^\gamma \varphi \Pi_{m=1}^{k-1}e^{i\pi S_m^\gamma}} e^{i\pi S_k^\alpha}, &\alpha\neq \gamma
\end{cases}.
\end{flalign}
Thus 
for $\alpha=\gamma$, Eq. (\ref{eq:double_expectation_a}) can be further simplified as
\begin{flalign}
&\langle\langle\sigma^\alpha_{N+1}\rangle\rangle=\nn\\
&\bra{G}
\big[\delta_{\alpha z}+\sigma_0^x (\delta_{\alpha x}+\delta_{\alpha y})\big]\big(\Pi_{j=1}^Ne^{i\pi S_j^\alpha}\big)\cdot\sigma^\alpha_{N+1}\ket{G};
\label{eq:double_expectation_b}
\end{flalign}
and for $\alpha\neq \gamma$, Eq. (\ref{eq:double_expectation_a}) can be simplified as
\bea
\langle\langle\sigma^\alpha_{N+1}\rangle\rangle&=&
\bra{G}
\big[\delta_{\alpha z}+\sigma_0^x (\delta_{\alpha x}+\delta_{\alpha y})\big]e^{iS_k^\gamma \varphi \Pi_{m=1}^{k-1}e^{i\pi S_m^\gamma}} \nn\\
&&\big(\Pi_{j=1}^Ne^{i\pi S_j^\alpha}\big)\cdot\sigma^\alpha_{N+1}\ket{G}.
\label{eq:double_expectation_c}
\eea
By expressing $e^{iS_k^\gamma \varphi \Pi_{m=1}^{k-1}e^{i\pi S_m^\gamma}} $ as
\begin{flalign}
e^{iS_k^\gamma \varphi \Pi_{m=1}^{k-1}e^{i\pi S_m^\gamma}} =\cos(S_k^\gamma \varphi)+i\sin(S_k^\gamma \varphi)\Pi_{m=1}^{k-1}e^{i\pi S_m^\gamma},
\end{flalign}
the $\alpha\neq \gamma$ case can be rewritten as
\begin{flalign}
&\langle\langle\sigma^\alpha_{N+1}\rangle\rangle=\nn\\
&\bra{G}\big[\delta_{\alpha z}+\sigma_0^x (\delta_{\alpha x}+\delta_{\alpha y})\big]\cos(S_k^\gamma \varphi)\big(\Pi_{j=1}^Ne^{i\pi S_j^\alpha}\big)\cdot\sigma^\alpha_{N+1}\ket{G}\nn\\
&+i\bra{G}\big[\delta_{\alpha z}+\sigma_0^x (\delta_{\alpha x}+\delta_{\alpha y})\big]\sin(S_k^\gamma \varphi)\big(\Pi_{m=1}^{k-1}e^{i\pi S_m^\gamma}\big)\nn\\
&\cdot\big(\Pi_{j=1}^Ne^{i\pi S_j^\alpha}\big)\cdot\sigma^\alpha_{N+1}\ket{G}.
\label{eq:double_expectation_d}
\end{flalign}

We take $\gamma=z$ as an example
(the analysis for $\gamma=x,y$ are similar). 
In this case, we have
\begin{flalign}
&\langle\langle\sigma^x_{N+1}\rangle\rangle=\bra{G}\cos(S_k^z \varphi)\sigma_0^x\big(\Pi_{j=1}^Ne^{i\pi S_j^x}\big)\sigma^x_{N+1}\ket{G}\nn\\
&+i\bra{G} \sin(S_k^z \varphi)\sigma_0^x\big(\Pi_{m=1}^{k-1}e^{i\pi S_m^z}\big)\big(\Pi_{j=1}^Ne^{i\pi S_j^x}\big)\sigma^x_{N+1}\ket{G}\nn\\
&\langle\langle\sigma^y_{N+1}\rangle\rangle=\bra{G}\cos(S_k^z \varphi)\sigma_0^x\big(\Pi_{j=1}^Ne^{i\pi S_j^y}\big)\sigma^y_{N+1}\ket{G}\nn\\
&+i\bra{G} \sin(S_k^z \varphi)\sigma_0^x\big(\Pi_{m=1}^{k-1}e^{i\pi S_m^z}\big)\big(\Pi_{j=1}^Ne^{i\pi S_j^y}\big)\sigma^y_{N+1}\ket{G}\nn\\
&\langle\langle\sigma^z_{N+1}\rangle\rangle=\bra{G}\big(\Pi_{j=1}^Ne^{i\pi S_j^z}\big)\sigma^z_{N+1}\ket{G}.
\label{eq:double_prime_expectation_a}
\end{flalign}
To proceed on, we state the following Lemma. 

\begin{Lemma}\label{lemma:vanish}
Let $C$ be an operator which anti-commutes with one of the symmetry operators $U_\alpha$ ($\alpha\in\{x,y,z\}$), where $U_\alpha$ is defined in Eq. (\ref{eq:sym}).
Then $\langle G|C|G \rangle =0$.
\end{Lemma}

{\em{Proof of Lemma~\ref{lemma:vanish}.}}
The proof is straightforward. 
Using $U_\alpha \ket{G}=\ket{G}$ and $\{C,U_\alpha\}=0$, we obtain $\bra{G}C\ket{G}=\bra{G}CU_\alpha\ket{G}=-\bra{G}U_{\alpha}C\ket{G}=-\bra{G}C\ket{G}=0$. ~~~$\Box$\medskip

It is straightforward to verify the following anti-commutation relations,
\begin{flalign}
&\{U_x, \sin(S_k^z \varphi)\sigma_0^x\big(\Pi_{m=1}^{k-1}e^{i\pi S_m^z}\big)\big(\Pi_{j=1}^Ne^{i\pi S_j^x}\big)\sigma^x_{N+1}\}=0\nn\\
&\{U_y,\cos(S_k^z \varphi)\sigma_0^x\big(\Pi_{j=1}^Ne^{i\pi S_j^y}\big)\sigma^y_{N+1}\}=0\nn\\
&\{U_x,\big(\Pi_{j=1}^Ne^{i\pi S_j^z}\big)\sigma^z_{N+1}\}=0.
\end{flalign}
Hence the expressions in Eq. (\ref{eq:double_prime_expectation_a}) reduce to
\begin{flalign}
&\langle\langle\sigma^x_{N+1}\rangle\rangle=\bra{G}\cos(S_k^z \varphi)\ket{G}\nn\\
&\langle\langle\sigma^y_{N+1}\rangle\rangle=\nn\\
&i\bra{G} \sin(S_k^z \varphi)\sigma_0^x\big(\Pi_{m=1}^{k-1}e^{i\pi S_m^x}\big)\big(\Pi_{j=k}^Ne^{i\pi S_j^y}\big)\sigma^y_{N+1}\ket{G}\nn\\
&\langle\langle\sigma^z_{N+1}\rangle\rangle=0.
\label{eq:double_prime_expectation_b}
\end{flalign}
Since $\sigma_0^x\big(\Pi_{m=1}^{k-1}e^{i\pi S_m^x}\big)\big(\Pi_{j=k}^Ne^{i\pi S_j^y}\big)\sigma^y_{N+1}$ can be written as
\begin{flalign}
&i\sigma_0^x\big(\Pi_{m=1}^{k-1}e^{i\pi S_m^x}\big)\big(\Pi_{j=k}^Ne^{i\pi S_j^y}\big)\sigma^y_{N+1}\nn\\
&=-U_x \big(\Pi_{j=k}^Ne^{i\pi S_j^z}\big)\sigma_{N+1}^z, 
\end{flalign}
$\{U_x,\sin(S_k^z\varphi)\}=0$, and $U_x\ket{G}=\ket{G}$, we obtain
\begin{flalign}
&\langle\langle\sigma^y_{N+1}\rangle\rangle=\bra{G} \sin(S_k^z\varphi) \big(\Pi_{j=k}^Ne^{i\pi S_j^z}\big)\sigma_{N+1}^z\ket{G}.
\label{eq:sigma_y_single_rot}
\end{flalign}
Combining Eq. (\ref{eq:double_prime_expectation_b}) and Eq. (\ref{eq:sigma_y_single_rot}), 
we obtain Eq. (\ref{eq:z_rotate_1}), thereby proving Proposition \ref{proposition:single_rot}.

%%%%%%%%%%%%%%%%%%%%%%%%%%%%%%%%%%%%%%%%%%%%%%%%%%%%%%
\section{Algebraic approach to SPT-MBQC}
\label{sec:algebraic_MBQC}

\begin{figure}%[htbp]
\begin{center}
\includegraphics[width=8.7cm]{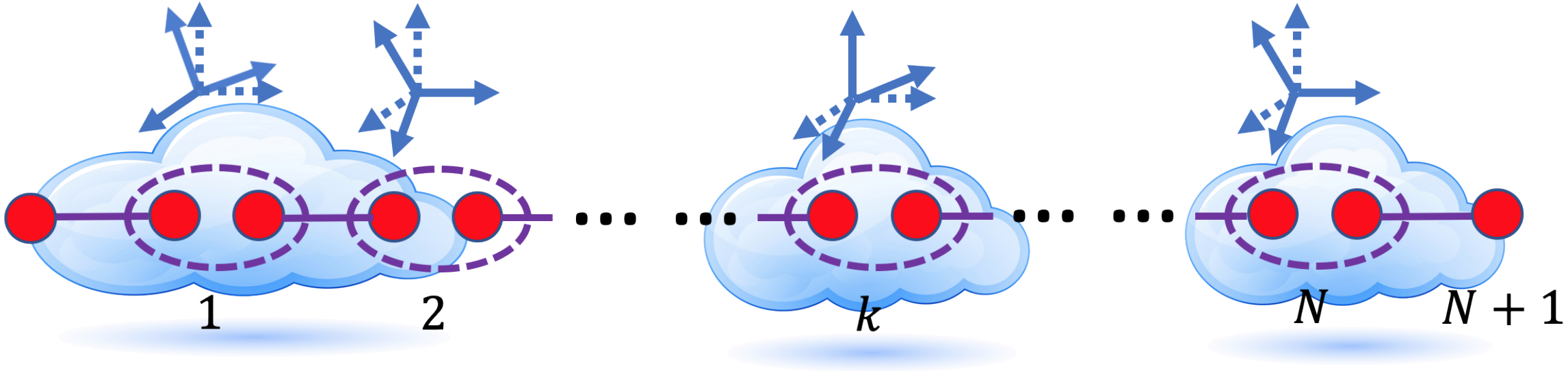}
\caption{Illustration of the measurement setting for realizing multiple logical rotations in SPT-MBQC. 
The spin-$1$ sites are in general measured in rotated basis. 
} \label{fig:multi_rot_SPT} 
\end{center}
\end{figure}

Motivated by the discussions in Sec. \ref{sec:SPT_MBQC_sum_path}, we formulate an algebraic approach of SPT-MBQC which applies to the general case of multiple rotation sites and arbitrary spin values, as illustrated in Fig. \ref{fig:multi_rot_SPT}.
The formalism in  this section is a generalization of Ref. \onlinecite{Raussendorf2023}.
The symmetry group is assumed to be $G=(\mathbb{Z}_2)^m$ as in Ref. \onlinecite{Raussendorf2023}.

%-------------------------------------------- 
%--------------------------------------------
 
%------------------------------------------------------------------------------------------------------------------------
\subsection{Representations}
\label{sec:representation}

We consider a finite chain of $N+2$ blocks, numbered from $0$ to $N+1$.

\subsubsection{Linear and projective representations}

For blocks $i=0,1,...,N+1$, let $\mathcal{H}_i$ be the Hilbert space of block $i$, and $\text{End}(\mathcal{H}_i)$  the set of linear operators on $\mathcal{H}_i$. 
We assume a number of linear and projective representations of $G=(\mathbb{Z}_2)^m$ on $\mathcal{H}_i$ and $\text{End}(\mathcal{H}_i)$.

\begin{enumerate}

\item{
At block $0$, we have linear representation $u_0$ on $\mathcal{H}_0$, and projective representation $v_{R,0}$ on $\mathcal{H}_0$.}

\item{
For blocks $1\leq i\leq N$, we have linear representation $u_i$ on $\mathcal{H}_i$.
Then $\text{End}(\mathcal{H}_i)$ naturally becomes a representation of $G$, where the conjugate  action $\varphi_i$ is defined as
\begin{flalign}
& \varphi_i(g^\prime) (A):=u_i(g^\prime) A [u_i(g^\prime)]^{-1},\nn\\
 &\forall g^\prime \in G, \forall A\in \text{End}(\mathcal{H}_i).
\end{flalign}

Since $G$ is an abelian group, any irreducible linear representation of $G$ is one-dimensional.
%Hence a basis of $\text{End}(\mathcal{H}_i)$ can be chosen such that each basis vector is a one-dimensional representation of $G$ via $\varphi_i$.
We assume that each $g\in G$ is associated with a linear representation $\mathcal{S}_i(g)\in \text{End}(\mathcal{H}_i)$ under $\varphi_i$. 
The corresponding character is denoted as $\chi_{i,g}$, i.e.,
\bea
\varphi_i(g^\prime) (\mathcal{S}_i(g))=\chi_{i,g}(g^\prime) \mathcal{S}_i(g),\forall g^\prime\in G.
\eea}
%Clearly, $\{\mathcal{S}_i(g)|g\in G\}$ is a subset of the above mentioned set of basis vectors.

\item{
At block $N+1$, we have projective representation $v_{L,N+1}$ on $\mathcal{H}_{N+1}$.}
\end{enumerate}

We further assume that the operators $u_i(g)$, $v_{R,0}(g)$, $v_{L,N+1}(g)$ and $\mathcal{S}_i(g)$ ($0\leq i\leq N,g\in G$) are all hermitian.

\subsubsection{Conditions for representations}
\label{subsec:conditions}

The above listed representations satisfy the following conditions. 

\begin{enumerate}
\item{The linear representation $u_0$ on block $0$ is special. 
Suppose $H$ is a maximal subgroup of the symmetry group $G$ satisfying
\begin{equation}\label{eq:u0}
[v_{R,0}(h),v_{R,0}(h')]=0,\; \forall h,h'\in H.
\end{equation}
Then for $H$ it holds that
\begin{equation}
\label{eq:H_spec}
u_0(h) = v_{R,0}(h),\;\forall h\in H.
\end{equation}}
\item{ 
The commutation relations among the elements of the projective representations $v_{R,0}$ and $v_{L,N+1}$ are given by
\begin{flalign}
\label{eq:KappaL}
 & \forall g,g'\in G,\nn\\
 &v_{R,0}(g) v_{R,0}(g') = (-1)^{\kappa(g,g')} v_{R,0}(g') v_{R,0}(g),\nn\\
&v_{L,N+1}(g) v_{L,N+1}(g') = (-1)^{\kappa(g,g')} v_{L,N+1}(g') v_{L,N+1}(g),
\end{flalign}
in which $\kappa$ is a function on the Cartesian product $G\times G$, namely, $\kappa: G\times G \longrightarrow \mathbb{Z}_2$. }
\item{
For each block $1\leq i\leq N$ in the bulk, there associates a subset of $G$, denoted as $\mathcal{G}_i$.
We require 
\bea
\chi_{i,g^\prime}=(-)^{\kappa(\cdot,g^\prime)},~ \forall g^\prime\in \mathcal{G}_i,
\eea 
namely,
\begin{flalign}
&\forall g\in G,\forall g^\prime \in \mathcal{G}_i,\nn\\
&u_i(g) \mathcal{S}_i(g^\prime) [u_i(g)]^{-1}=(-)^{\kappa(g,g^\prime)} \mathcal{S}_i(g^\prime).
\label{eq:condition_Sg}
\end{flalign}}
\item{The linear representation $U$ of $G$ on the entire spin chain is given by
\begin{equation}
U(g)=v_{R,0}(g)\left(\bigotimes_{i=1}^N u_i(g)\right) v_{L,N+1}(g),\;\; \forall g\in G,
\label{eq:Udef}
\end{equation}
which are symmetries of the system. }
\end{enumerate}

\subsubsection{MBQC resource state}

The short-range entangled many-body state $\ket{\Psi}$ on the chain which serves as the MBQC resource state is a symmetric state under the $(\mathbb{Z}_2)^m$ symmetry group, satisfying
\bea
U(g)\ket{\Psi}=(-)^{\chi(g)} \ket{\Psi},~\forall g\in G,
\label{eq:Psi_sym}
\eea
in which $\chi(g)\in\{0,1\}$. 

%------------------------------------------------------------------------------------------------------------------------
\subsection{Logical observables}

\subsubsection{Logical observables}

The encoded Pauli operators are constructed  as
\bea
\overline{T}(g)=\left(\bigotimes_{i=0}^{N} u_i(g) \right)v_{L,N+1}(g). 
\label{eq:def_Tbar}
\eea
%Denote $\mathcal{L}$ to be the linear space spanned by $\overline{T}(g)$'s with coefficients in $\mathbb{C}$, i.e.,
%\bea
%\mathcal{L}=\text{Span} \{\overline{T}(g)~|~g\in G\}.
%\label{eq:Lie_space}
%\eea

The expectation values of the symmetric state $\ket{\Psi}$ under logical observables are given by the following Lemma.

\begin{Lemma}\label{lemma:L_init}
The symmetric state $\ket{\Psi}$ in Eq. (\ref{eq:Psi_sym}) satisfies
\begin{equation}\label{eq:InitEval}
\begin{array}{rcll}
\bra{\Psi} \overline{T}(g)\ket{\Psi} &=& \displaystyle{(-1)^{\chi(g)}},& \text{if } g \in H,\\
\bra{\Psi} \overline{T}(g)\ket{\Psi} &=& 0,& \text{if } g \in G\backslash H.
\end{array}
\end{equation}
\end{Lemma}

{\em{Proof of Lemma~\ref{lemma:L_init}.}}
We refer to Lemma 5 in Ref. \onlinecite{Raussendorf2023} for the proof.  ~~~$\Box$\medskip

%------------------------------------------------------------------------------------------------------------------------
\subsubsection{Logical space}

Define a subspace $\mathcal{Q}$ as 
\bea
\mathcal{Q}=\text{Span} \{\overline{T}(g)\ket{\Psi}~|~g\in G\}.
\label{eq:logical_space}
\eea
Denote $\mathcal{P}$ to be the projection operator onto $\mathcal{Q}$.

\begin{Lemma}\label{lemma:irred}
$\mathcal{Q}$ is a finite-dimensional irreducible representation of the group generated by $\{T(g)~|~ g\in G\}$. 
\end{Lemma}

{\em{Proof of Lemma~\ref{lemma:irred}.}}
The dimension of $\mathcal{Q}$ defined in Eq. (\ref{eq:logical_space}) is bounded from above by the order of the group $G$,
hence $\mathcal{Q}$ is finite-dimensional. 

Since $v_{L,N+1}$ is a projective representation of $G=(\mathbb{Z}_2)^m$, we have
\bea
v_{L,N+1}(g)v_{L,N+1}(g^\prime)=e^{i\varphi(g,g^\prime)} v_{L,N+1}(gg^\prime),
\eea
in which $e^{i\varphi(g,g^\prime)}$ is a phase factor depending on $g$ and $g^\prime$. 
Using Eq. (\ref{eq:def_Tbar}), it is straightforward to see that 
\bea
\overline{T}(g)\overline{T}(g^\prime) = e^{i\varphi(g,g^\prime)} \overline{T}(gg^\prime).
\eea
Let $\ket{\psi}$ be any state in $\mathcal{Q}$ given by
\bea
\ket{\psi}=\sum_{g\in G}\alpha_g \overline{T}(g)\ket{\Psi},
\eea
in which $\alpha_g\in \mathbb{C}$.
Then for any $g^\prime \in G$, we have
\bea
\overline{T}(g^\prime)\ket{\psi}=\sum_{g\in G}\alpha_g e^{i\varphi(g^\prime,g)} \overline{T}(g^\prime g)\ket{\Psi}\in \mathcal{Q}.
\eea
This shows that $\mathcal{Q}$ is a representation of the group $\langle \{\overline{T}(g)~|~ g\in G\} \rangle $,
where $\langle \{...\} \rangle$ denotes the group generated by the elements in the set $\{...\}$. 

Define $\mathcal{F}(H)$ to be the collection of all $\mathbb{Z}_2$-valued functions on $H$ (i.e., valued in $\{0,1\}$), where $H$ is the maximal subgroup of $G$ introduced in Sec. \ref{subsec:conditions}. 
Since $\overline{T}(h)$'s mutually commute with each other for all $h\in H$,
$\mathcal{Q}$ can be decomposed into a direct sum of common eigenspaces of  the operators in $\{\overline{T}(h)~|~h\in H\}$. 
Notice that since $[\overline{T}(g)]^2$ for any $g\in G$ can be made equal to the identity operator by adjusting the phase factors of $\overline{T}(g)$  (for a proof, see Lemma 4 in Ref. \onlinecite{Raussendorf2023}),
the eigenvalues of $\overline{T}(h)$'s are $\pm 1$. 
Therefore, any common eigenspace (or common eigenvector) $\mathcal{E}$ of $\{\overline{T}(h)~|~h\in H\}$  induces $\lambda_{\mathcal{E}}\in \mathcal{F}(H)$, where $(-)^{\lambda_{\mathcal{E}}(h)}$ is the eigenvalue of $\overline{T}(h)$ for $\mathcal{E}$.
We call $\lambda_{\mathcal{E}}$ an eigenvalue function on $H$.
In particular, $\lambda_{\ket{\Psi}}(h)=\chi(h)$, where $\chi(h)$ is given by Eq. (\ref{eq:InitEval}). 

Denote $\mathcal{Q}_{\lambda_{\ket{\Psi}}}$ to be the common eigenspace of $\{\overline{T}(h)~|~h\in H\}$ in $\mathcal{Q}$ having eigenvalue function $\lambda_{\ket{\Psi}}$.
Next we show that $\mathcal{Q}_{\lambda_{\ket{\Psi}}}$ is one-dimensional. 
For each coset $gH\in G/H$, we choose a representative element $\tilde{g}\in gH$.
When $gH=H$, the representative element is chosen to be the identity element $e$.
For any element $g\in G$, there exist $\tilde{g}$ and $h\in H$, such that $g=\tilde{g}h$.
Consider a basis state $\overline{T}(g)\ket{\Psi}\in \mathcal{Q}$. 
Since
\bea
\overline{T}(g)\ket{\Psi}&=&\overline{T}(\tilde{g}h)\ket{\Psi}\nn\\
&=&e^{-i \varphi(\tilde{g},h)} \overline{T}(\tilde{g})\overline{T}(h)\ket{\Psi}\nn\\
&=&(-)^{\chi(h)}e^{-i \varphi(\tilde{g},h)} \overline{T}(\tilde{g})\ket{\Psi},
\eea
we see that $\{\overline{T}(\tilde{g})\ket{\Psi}~|~\tilde{g}\}$ constitute a set of basis states for $\mathcal{Q}$.
When $g\not\in H$ (i.e., $\tilde{g}\neq e$), there exists $f\in H$, such that $\{\overline{T}(\tilde{g}),\overline{T}(f)\}=0$,
otherwise $\overline{T}(\tilde{g})$ commutes with all elements in $H$, contradicting the maximal property of $H$ assumed in Sec. \ref{subsec:conditions}.
As a result, the eigenvalues of $\overline{T}(f)$ for $\overline{T}(\tilde{g})\ket{\Psi}$ and $\ket{\Psi}$ differ by a sign.
This shows that $\lambda_{\overline{T}(\tilde{g})\ket{\Psi}}\neq \lambda_{\ket{\Psi}}$ when $\tilde{g}\neq e$, since $\lambda_{\overline{T}(\tilde{g})\ket{\Psi}}(f)\neq \lambda_{\ket{\Psi}}(f)$.
Hence, $\ket{\Psi}$ is the only state among the basis states $\{\overline{T}(\tilde{g})\ket{\Psi}~|~ {\tilde{g}}\}$ having eigenvalue function $\lambda_{\ket{\Psi}}$.
Namely, the dimension of $\mathcal{Q}_{\lambda_{\ket{\Psi}}}$ is one. 

Now we are prepared to show that $\mathcal{Q}$ is irreducible. 
Suppose on the contrary $\mathcal{Q}$ is not an irreducible representation of $\langle \{\overline{T}(g)~|~ g\in G\} \rangle $. 
According to Maschke's theorem,  any representation of a finite  group can be decomposed into a direct sum of irreducible ones \cite{Serre1977}. 
Hence $\mathcal{Q}$ can be decomposed as $\mathcal{Q}_1\oplus \mathcal{Q}_2$, where both $\mathcal{Q}_1$ and $\mathcal{Q}_2$ are representations (not necessarily irreducible) of the group $\langle \{\overline{T}(g)~|~ g\in G\} \rangle $. 
Each $\mathcal{Q}_i$ ($i=1,2$) can be decomposed into a direct sum of common eigenspaces of $\{\overline{T}(h)~|~h\in H\}$. 
Since $\mathcal{Q}_{\lambda_{\ket{\Psi}}}$ is one-dimensional, it is contained in either $\mathcal{Q}_1$ or $\mathcal{Q}_2$ (otherwise if $\ket{\Psi}$ has components in both $\mathcal{Q}_1$ and $\mathcal{Q}_2$, $\mathcal{Q}_{\lambda_{\ket{\Psi}}}$ will be at least two-dimensional in $\mathcal{Q}$).
Suppose $\mathcal{Q}_{\lambda_{\ket{\Psi}}}$ is in $\mathcal{Q}_1$,
then $\ket{\Psi}\in \mathcal{Q}_1$.
Since by assumption $\mathcal{Q}_1$ is a representation of the group $\langle \{\overline{T}(g)~|~ g\in G\} \rangle $,
$\overline{T}(g)\ket{\Psi}$ is in $\mathcal{Q}_1$ which holds for any $g\in G$.
This means $\mathcal{Q}\subseteq\mathcal{Q}_1$, i.e., $\mathcal{Q}=\mathcal{Q}_1$, contradicting the assumption $\mathcal{Q}=\mathcal{Q}_1\oplus \mathcal{Q}_2$.
Therefore, the assumption of $\mathcal{Q}$ being reducible does not hold, hence $\mathcal{Q}$ must be an irreducible representation.
~~~$\Box$\medskip

%------------------------------------------------------------------------------------------------------------------------
\subsubsection{Evolved logical observables}

Define $L_k(g)$ as ($ g\in \mathcal{G}_k,1\leq k\leq N$)
\bea
L_k(g)&=&\left(\bigotimes_{i=0}^{k-1} u_i(g) \right)\mathcal{S}_k(g).
\eea
The unitary gates $V_k$ and the accumulated gates $V_{\leq k}$ are defined as
\bea
V_k&=&\text{exp}\left(-i\frac{\alpha_k}{2}L_k(g_k)\right),\nn\\
V_{\leq k}&=&V_kV_{k-1}...V_2V_1.
\label{eq:Vdef}
\eea
Using $V_{\leq k}$, the evolved logical observables $\{\overline{T}_t(g)\}_{g\in G}$ are defined as
\bea
\overline{T}_t(g)&=&V_{\leq t}^\dagger\overline{T}(g)V_{\leq t}. 
\label{eq:GateSeqs}
\eea
%Let $k=n$ in Eq. (\ref{eq:GateSeqs}), we have $\overline{T}_n=V_{\leq n}^\dagger \overline{T}(g)V_{\leq n}$.

We further define $R_k(g,\alpha_k)$ as
\begin{flalign}
%R_k(g)&=&\mathcal{S}_k(g)u_k(g) \left(\bigotimes_{i=k+1}^{n} u_i(g) \right) v_{L,n+1}(g),\nn\\
R_k(g,\alpha_k)&=&\frac{\sin\left[\mathcal{S}_k(g)\alpha_k\right]}{\sin(\beta_k(\alpha_k))}u_k(g) \left(\bigotimes_{i=k+1}^{N} u_i(g) \right) v_{L,n+1}(g),
\label{eq:string_def}
\end{flalign}
where $\beta_k(\alpha_k)$ is defined by 
\bea
\beta_k(\alpha_k)=\arccos\left[ \langle\Psi| \cos(\mathcal{S}_k(g)\alpha_k)|\Psi\rangle\right].
\label{eq:Def_theta_k}
\eea
%Notice that unlike Ref. \cite{Raussendorf2022}, the rotation angle in $V_k$ is $\theta_k$, not $\alpha_k$.
%in which $\theta_k$ is defined in Eq. (\ref{eq:Def_theta_k}).
The string order parameter $\sigma_k(g,\alpha_k)$ is  the expectation value of $R_k(g,\alpha_k)$ as
\bea
\sigma_k(g,\alpha_k)=\langle\Psi| R_k(g,\alpha_k)|\Psi\rangle.
\eea

Notice that  unlike Ref. \onlinecite{Raussendorf2023}, the string order parameter now in general depends on the angle $\alpha_k$.
On the other hand, in the limit $\alpha_k\ll 1$, $R_k(g,\alpha_k)$ approaches $\nu_k(g)\alpha_k$, where $\nu_k(g)$ is given by
\bea
\nu_k(g)=\mathcal{S}_k(g)u_k(g) \left(\bigotimes_{i=k+1}^{N} u_i(g) \right) v_{L,n+1}(g),
\label{eq:string_nu}
\eea
which is a bulk-to-end string order operator, independent of the angle $\alpha_k$.

There is a useful Lemma which gives the commutation relations between $\cos(\mathcal{S}_k(g^\prime)\alpha_k)$,  $L_k(g)$, $R_k(g,\alpha_k)$ and the logical observables. 

\begin{Lemma}\label{lemma:cd}It holds that $\forall g\in G, g'\in {\cal{G}}_k,1\leq k\leq N$,
\begin{flalign}
&[\cos(\mathcal{S}_k(g^\prime)\alpha_k),\overline{T}(g)]=0,\label{eq:Sk_g_comm}\\
&[R_k(g'),T(g)]=0,\label{eq:R_comm} \\
&[L_k(g'),\overline{T}(g)]=0 \Longleftrightarrow [\overline{T}(g'),\overline{T}(g)]=0,\label{eq:L_comm}\\
&\{L_k(g'),\overline{T}(g)\}=0 \Longleftrightarrow \{\overline{T}(g'),\overline{T}(g)\}=0. \label{eq:L_acomm}
\end{flalign}
\end{Lemma}

{\em{Proof of Lemma~\ref{lemma:cd}.}}
Note: this Lemma is the counterpart of Lemma $4$ in Ref. \onlinecite{Raussendorf2023}. 

Eq. (\ref{eq:Sk_g_comm}) can be proved as follows
\bea
&&\overline{T}(g)\cos\left(\mathcal{S}_k(g^\prime)\alpha_k\right)[\overline{T}(g)]^{-1}\nn\\
&=&u_k(g)\cos\left(\mathcal{S}_k(g^\prime)\alpha_k\right)[u_k(g)]^{-1}\nn\\
&=&\cos\left(u_k(g)\mathcal{S}_k(g^\prime)[u_k(g)]^{-1}\alpha_k\right)\nn\\
&=&\cos\left((-)^{\kappa(g,g^\prime)}\mathcal{S}_k(g^\prime)\alpha_k\right)\nn\\
&=&\cos\left(\mathcal{S}_k(g^\prime)\alpha_k\right),
\eea
in which Eq. (\ref{eq:condition_Sg}) is used.

For Eq. (\ref{eq:R_comm}), using 
\bea
&&u_k(g)\sin\left[\mathcal{S}_k(g^\prime)\alpha_k\right][u_k(g)]^{-1}\nn\\
&=&\sin\left(u_k(g)\mathcal{S}_k(g^\prime)[u_k(g)]^{-1}\alpha_k\right),
\eea
we obtain (for any $g\in G,g^\prime\in \mathcal{G}_k$)
\bea
%\begin{array}{rcl}
&&\overline{T}(g)R_k(g',\alpha_k)[\overline{T}(g)]^{-1} \nn\\
&=&  (-1)^{\kappa(g,g')+\kappa(g,g')}  R_k(g',\alpha_k) \nn\\
&=&R_k(g',\alpha_k),
%\end{array}
\eea
in which Eq. (\ref{eq:KappaL}) is used at block $N+1$, Eq. (\ref{eq:condition_Sg}) is used at block $k$, and $u_i$ being linear representations is used at other blocks.

For Eq. (\ref{eq:L_comm}) (for any $g\in G,g^\prime\in \mathcal{G}_k$),
\bea
&&[L_k(g'),\overline{T}(g)]=0\nn\\
&\Longleftrightarrow&\overline{T}(g)L_k(g')[\overline{T}(g)]^{-1}=L_k(g')\nn\\
&\Longleftrightarrow& u_k(g)\mathcal{S}_k(g^\prime)[u_k(g)]^{-1}=\mathcal{S}_k(g^\prime)\nn\\
&\Longleftrightarrow& (-)^{\kappa(g,g^\prime)}=1\nn\\
&\Longleftrightarrow& [v_{L,N+1}(g),v_{L,N+1}(g^\prime)]=0\nn\\
&\Longleftrightarrow& [\overline{T}(g'),\overline{T}(g)]=0.
\eea

For Eq. (\ref{eq:L_acomm}) (for any $g\in G,g^\prime\in \mathcal{G}_k$),
\bea
&&\{L_k(g'),\overline{T}(g)\}=0\nn\\
&\Longleftrightarrow& \overline{T}(g)L_k(g')[\overline{T}(g)]^{-1}=-L_k(g')\nn\\
&\Longleftrightarrow& u_k(g)\mathcal{S}_k(g^\prime)[u_k(g)]^{-1}=-\mathcal{S}_k(g^\prime)\nn\\
&\Longleftrightarrow& (-)^{\kappa(g,g^\prime)}=-1\nn\\
&\Longleftrightarrow& \{v_{L,N+1}(g),v_{L,N+1}(g^\prime)\}=0\nn\\
&\Longleftrightarrow& \{\overline{T}(g'),\overline{T}(g)\}=0.
\eea 
This completes the proof of the Lemma. ~~~$\Box$\medskip

Define an operation $\mathcal{M}_k(g,\alpha_k)$ acting on $\overline{T}(g^\prime)$ ($g\in\mathcal{G}_k$, $g^\prime\in G$) as 
%\begin{widetext}
\begin{flalign}
&\mathcal{M}_k(g,\alpha_k)(\overline{T}(g^\prime))=
\overline{T}(g^\prime),\text{ if }\kappa(g,g^\prime)= 0
\label{eq:Mcal_commu}
\end{flalign}
and
\begin{flalign}
&\mathcal{M}_k(g,\alpha_k)(\overline{T}(g^\prime))=\cos(\mathcal{S}_k\alpha_k) \overline{T}(g^\prime)\nn\\
&-\sin(\beta_k(\alpha_k)) R_k(g,\alpha_k) i \overline{T}(g)\overline{T}(g^\prime),\text{ if }\kappa(g,g^\prime)\neq 0. 
\label{eq:Mcal}
\end{flalign}
%\end{widetext}
Notice that $i \overline{T}(g)\overline{T}(g^\prime)$ is just $ \overline{T}(gg^\prime)$ up to a phase. 
By arranging $\overline{T}(g)$'s in a column vector, 
the action of $\mathcal{M}_k$ can be written in the following matrix form
\bea
\left(\begin{array}{c}
\mathcal{M}_k(g,\alpha_k)(\overline{T}(g_1))\\
\mathcal{M}_k(g,\alpha_k)(\overline{T}(g_2))\\
:\\
\mathcal{M}_k(g,\alpha_k)(\overline{T}(g_{|G|}))\\
\end{array}\right)
=M_k(g,\alpha_k)
\left(\begin{array}{c}
\overline{T}(g_1)\\
\overline{T}(g_2)\\
:\\
\overline{T}(g_{|G|})\\
\end{array}\right),
\label{eq:Mcal_matrix_form}
\eea
in which the matrix elements of $|G|\times |G|$ matrix $M_k(g,\alpha_k)$ are operators on the Hilbert space that can be read from Eq. (\ref{eq:Mcal_commu}) and Eq. (\ref{eq:Mcal}). 
Notice that according to Lemma \ref{lemma:cd}, the matrix elements of $M_k(g,\alpha_k)$ commute with $\overline{T}(g)$ ($\forall g\in G$).

We have the following Lemma for $\overline{T}_t(g)$.

\begin{Lemma}\label{lemma:overline_T_t}
$\{\overline{T}_t(g)~|~g\in G\}$ are related to $\{\overline{T}(g)~|~g\in G\}$ by
\bea
\left(\begin{array}{c}
\overline{T}_t(g_1)\\
\overline{T}_t(g_2)\\
:\\
\overline{T}_t(g_{|G|})\\
\end{array}\right)
=
\Pi_{k=1}^t M_k(g_k,\alpha_k)
\left(\begin{array}{c}
\overline{T}(g_1)\\
\overline{T}(g_2)\\
:\\
\overline{T}(g_{|G|})\\
\end{array}\right),
\label{eq:overline_T_t}
\eea
in which $\Pi_{k=1}^t M_k=M_tM_{t-1}...M_1$.
\end{Lemma}

{\em{Proof of Lemma~\ref{lemma:overline_T_t}.}}
We will prove the following relation
\bea
\left(\begin{array}{c}
\overline{T}_{k}(g_1)\\
\overline{T}_{k}(g_2)\\
:\\
\overline{T}_{k}(g_{|G|})\\
\end{array}\right)
=
M_k(g_k,\alpha_k)
\left(\begin{array}{c}
\overline{T}_{k-1}(g_1)\\
\overline{T}_{k-1}(g_2)\\
:\\
\overline{T}_{k-1}(g_{|G|})\\
\end{array}\right).
\label{eq:T_recursive}
\eea
Then Lemma \ref{lemma:overline_T_t} follows from induction. 

Let's consider $V_k^\dagger\overline{T}(g)V_k$, where $V_k$ is defined in Eq. (\ref{eq:Vdef}). 
We distinguish two cases.

Case 1. $[\overline{T}(g_k),\overline{T}(g)]=0$

In this case, according to Lemma \ref{lemma:cd}, we have $[L_k(g_k),\overline{T}(g)]=0$.
Then 
\bea
V_k^\dagger\overline{T}(g)V_k&=&e^{i\frac{\alpha_k}{2}L_k(g_k)}\overline{T}(g)e^{-i\frac{\alpha_k}{2}L_k(g_k)}=\overline{T}(g).
\label{eq:Vk_T_a}
\eea

Case 2. $\{\overline{T}(g_k),\overline{T}(g)\}=0$

According to Lemma \ref{lemma:cd}, now we have $\{L_k(g_k),\overline{T}(g)\}=0$.
Then 
\bea
V_k^\dagger\overline{T}(g)V_k&=&e^{i\frac{\alpha_k}{2}L_k(g_k)}\overline{T}(g)e^{-i\frac{\alpha_k}{2}L_k(g_k)}\nn\\
&=&e^{i\alpha_k L_k(g_k)}\overline{T}(g)\nn\\
&=&[\cos\big(\alpha_k L_k(g_k)\big)+i\sin\big(\alpha_k L_k(g_k)\big)]\overline{T}(g)\nn\\
&=&\cos\big(\alpha_k \mathcal{S}_k(g_k)\big)\overline{T}(g)\nn\\
&&+i\sin\big(\alpha_k \mathcal{S}_k(g_k)\big)\left(\bigotimes_{i=0}^{k-1} u_i(g_k) \right)\overline{T}(g),
\eea
in which $[u_i(g)]^2=1$ is used. 
Using 
\bea
\bigotimes_{i=0}^{k-1} u_i(g_k)=\left(\bigotimes_{i=k}^{N}u_i(g)\right) v_{L,N+1}(g_k)\overline{T}(g_k), 
\eea
we obtain
\begin{flalign}
&V_k^\dagger\overline{T}(g)V_k=\cos\big(\alpha_k \mathcal{S}_k(g_k)\big) \overline{T}(g)\nn\\
&+i\sin\big(\alpha_k \mathcal{S}_k(g_k)\big)\left(\bigotimes_{i=k}^{N}u_i(g)\right) v_{L,N+1}(g_k)\overline{T}(g_k)\overline{T}(g).
\label{eq:evolved_acomm}
\end{flalign}

Combining Eq. (\ref{eq:Vk_T_a},\ref{eq:evolved_acomm}), and comparing with Eqs. (\ref{eq:Mcal_commu},\ref{eq:Mcal}), we see that 
\bea
V_k^\dagger\overline{T}(g)V_k= \mathcal{M}_k(g_k,\alpha_k)(\overline{T}(g^\prime)).
\eea
As a result,
\bea
V_{\leq k}^\dagger\overline{T}(g)V_{\leq k}= V_{\leq k-1}^\dagger\mathcal{M}_k(g_k,\alpha_k)(\overline{T}(g^\prime))V_{\leq k-1}.
\label{eq:VM}
\eea
Notice that both $\cos(\mathcal{S}_k\alpha_k)$ and $R_k(g_k,\alpha_k)$ commute with $V_{k-1}$,
since the supports of $\cos(\mathcal{S}_k\alpha_k)$ and $R_k(g_k,\alpha_k)$ are on sites $\{k\}$ and $\{j~|~j\geq k\}$, respectively, 
whereas the support of the operator $V_{k-1}$ is located on sites $\{j~|~j\leq k-1\}$.
Therefore, Eq. (\ref{eq:VM}) implies
\bea
\overline{T}_k(g)= \mathcal{M}_k(g_k,\alpha_k)(\overline{T}_{k-1}(g^\prime)).
\label{eq:VM_b}
\eea
Writing Eq. (\ref{eq:VM_b}) in a matrix form using the definition of $\mathcal{M}_k(g_k,\alpha_k)$ in Eq. (\ref{eq:Mcal_matrix_form}), we obtain Eq. (\ref{eq:T_recursive}).  ~~~$\Box$\medskip
 
%------------------------------------------------------------------------------------------------------------------------
\subsubsection{Projected evolved observables}

The projected evolved logical observables $\overline{T}^{(P)}_t(g)$ on the logical space are defined as
\bea
\overline{T}^{(P)}_t(g)=\mathcal{P}\overline{T}_t(g)\mathcal{P}. 
\eea

\begin{Lemma}\label{lemma:scalar}
Let $D$ be an operator satisfying $[D,\overline{T}(g)]=0$ for any $g\in G$. Then

\begin{enumerate}
\item{$\mathcal{P}D\mathcal{P}=\bra{\Psi} D\ket{\Psi}\mathcal{P}$,
\label{item:scalar}
}

\item{$\mathcal{P}D\overline{T}(g)\mathcal{P}=\bra{\Psi} D\ket{\Psi} \overline{T}^{(P)}(g)$,  $\forall g\in G$.
\label{item:factorize_P}
}
\end{enumerate}
\end{Lemma}

{\em{Proof of Lemma~\ref{lemma:scalar}.}}
We first prove Assertion \ref{item:scalar} in the Lemma.
From $[D,\overline{T}(g)]=0$, we know $[\mathcal{P}D\mathcal{P},\overline{T}(g)]=0$.
According to Lemma \ref{lemma:irred}, $\mathcal{Q}$ is an irreducible representation of the group generated by $\overline{T}(g)$'s ($g\in G$),
hence the operator $\mathcal{P}D\mathcal{P}$ must be proportional to the identity operator $\mathcal{P}$ in the space $\mathcal{Q}$ according to Schur's Lemma.
Writing $\mathcal{P}D\mathcal{P}=\lambda \mathcal{P}$, the scalar factor $\lambda$ clearly can be determined by taking the expectation value of $D$ over any state in $\mathcal{Q}$.
Choosing the state to be $\ket{\Psi}$, we arrive at Assertion \ref{item:scalar}.

Assertion \ref{item:factorize_P} follows directly from $[\overline{T}(g),\mathcal{P}]=0$ and Assertion \ref{item:scalar}. ~~~$\Box$\medskip

\begin{Lemma}\label{lemma:P_T}
The projected evolved operators satisfy
\begin{flalign}
\left(\begin{array}{c}
\overline{T}^{(P)}_t(g_1)\\
\overline{T}^{(P)}_t(g_2)\\
:\\
\overline{T}^{(P)}_t(g_{|G|})\\
\end{array}\right)
=
\bra{\Psi}\Pi_{k=1}^t M_k(g_k,\alpha_k)\ket{\Psi}
\left(\begin{array}{c}
\overline{T}^{(P)}(g_1)\\
\overline{T}^{(P)}(g_2)\\
:\\
\overline{T}^{(P)}(g_{|G|})\\
\end{array}\right).
\label{eq:overline_T_t_P}
\end{flalign}
\end{Lemma}

{\em{Proof of Lemma~\ref{lemma:P_T}.}}
According to Lemma \ref{lemma:cd},
 $M_k(g_k,\alpha_k)$ commutes with $\overline{T}(g)$.
Applying $\mathcal{P}$ to both sides of Eq. (\ref{eq:overline_T_t}) and using Lemma \ref{lemma:scalar}, we obtain Eq. (\ref{eq:overline_T_t_P}).  ~~~$\Box$\medskip

Taking expectation value over $\ket{\Psi}$ on both sides of Eq. (\ref{eq:overline_T_t_P}), we obtain
\begin{flalign}
&\left(\begin{array}{c}
\bra{\Psi}\overline{T}_t(g_1)\ket{\Psi}\\
\bra{\Psi}\overline{T}_t(g_2)\ket{\Psi}\\
:\\
\bra{\Psi}\overline{T}_t(g_{|G|})\ket{\Psi}\\
\end{array}\right)\nn\\
&=
\bra{\Psi}\Pi_{k=1}^t M_k(g_k,\alpha_k)\ket{\Psi}
\left(\begin{array}{c}
\bra{\Psi}\overline{T}(g_1)\ket{\Psi}\\
\bra{\Psi}\overline{T}(g_2)\ket{\Psi}\\
:\\
\bra{\Psi}\overline{T}(g_{|G|})\ket{\Psi}\\
\end{array}\right),
\label{eq:overline_T_t_Psi}
\end{flalign}
in which the column vector on the right hand side of Eq. (\ref{eq:overline_T_t_Psi}) is determined by Eq. (\ref{eq:InitEval}). 

There is a useful factorization property for products of bulk-to-end string order operators when their starting sites are separated by distances much larger than the entanglement length (for a definition of entanglement length, see Definition 1 in Ref. \onlinecite{Raussendorf2023}). 

\begin{Lemma}\label{lemma:sting_factorize}
Let $\ket{\Psi}$ be the symmetric state satisfying Eq. (\ref{eq:Psi_sym}).
Suppose $\ket{\Psi}$ is short-range entangled which can be created from a product state using a quantum circuit $W_\Psi$, where the circuit $W_\Psi$ has an entanglement range $\xi$. 
Let $\{E_{k_i}~|~1\leq i\leq m\}$ and $\{E_{l_j}~|~1\leq j\leq n\}$ be two sets of operators, 
where $E_k$ is either $\cos\big(\alpha_k \mathcal{S}_k(g_k)\big)$ or $\sin(\beta_k(\alpha_k)) R_k(g_k,\alpha_k)$,
and $l_j-k_i\gg \xi$, $\forall 1\leq i\leq m, 1\leq  j\leq n$.
Then
\begin{flalign}
\bra{\Psi}\Pi_{i=1}^m E_{k_i} \cdot \Pi_{j=1}^n E_{l_j}\ket{\Psi}
=\bra{\Psi}\Pi_{i=1}^m E_{k_i}\ket{\Psi}
\bra{\Psi} \Pi_{j=1}^n E_{l_j}\ket{\Psi}.
\label{eq:factorize_string}
\end{flalign}
\end{Lemma}

{\em{Proof of Lemma~\ref{lemma:sting_factorize}.}}
Let $h_1,...,h_p\in G$ be the $g_k$'s that appear in $R_k(g_k,\alpha_k)$'s in $\{E_{k_i}~|~1\leq i\leq m\}$. 
Then the operator on site $q$ ($k_m+1\leq q\leq l_1-1$) is $\Pi_{i=1}^p u_q(h_i)=u_q(\Pi_{i=1}^q h_i)$. 
Since $\overline{T}(\Pi_{i=1}^q h_i)$ is a symmetry of the state $\ket{\Psi}$, we have
\bea
&&\bra{\Psi}\Pi_{i=1}^m E_{k_i} \cdot \Pi_{j=1}^n E_{l_j}\ket{\Psi}\nn\\
&=&(-)^{\chi(\Pi_{i=1}^q h_i)}\bra{\Psi} \overline{T}(\Pi_{i=1}^q h_i)\Pi_{i=1}^m E_{k_i} \cdot \Pi_{j=1}^n E_{l_j}\ket{\Psi}.\nn\\
\label{eq:factorize_string2}
\eea
Notice that the operator on site $q$ ($k_m+1\leq q\leq l_1-1$) now becomes identity,
which means that $\overline{T}(\Pi_{i=1}^q h_i)\Pi_{i=1}^m E_{k_i} \cdot \Pi_{j=1}^n E_{l_j}$ breaks up into two pieces separated by a distance much larger than the characteristic entanglement length. 
Since the state $\ket{\Psi}$ is short-range entangled, the expectation value of these two far-separated pieces factorizes (for a proof, see Lemma 1 in Ref. \onlinecite{Raussendorf2023}), and we obtain
\bea
&&\bra{\Psi}\Pi_{i=1}^m E_{k_i} \cdot \Pi_{j=1}^n E_{l_j}\ket{\Psi}\nn\\
&=&(-)^{\chi(\Pi_{i=1}^q h_i)}\bra{\Psi} \overline{T}(\Pi_{i=1}^q h_i)\Pi_{i=1}^m E_{k_i} \ket{\Psi}\bra{\Psi} \Pi_{j=1}^n E_{l_j}\ket{\Psi}\nn\\
&=&\bra{\Psi}\Pi_{i=1}^m E_{k_i} \ket{\Psi}\bra{\Psi} \Pi_{j=1}^n E_{l_j}\ket{\Psi},
\label{eq:factorize_string3}
\eea
which proves Eq. (\ref{eq:factorize_string}).  ~~~$\Box$\medskip

We can distinguish two distinct regimes, the uncorrelated and correlated regimes.
For a given quantum algorithm, let $k_i$ ($1\leq i\leq r_t$, $k_1<k_2<...<k_{r_t}$) be the positions of the sites in $\Pi_{k=1}^t M_k(g_k,\alpha_k)$ where $\alpha_k\neq 0$. 
Then we have 
\bea
\Pi_{k=1}^t M_k(g_k,\alpha_k)=\Pi_{i=1}^{r_t} M_k(g_{k_i},\alpha_{k_i}),
\label{eq:_product_M_rt}
\eea
since $M_k(g_k,0)$ is the identity matrix as can be seen from Eq. (\ref{eq:Mcal_commu}) and Eq. (\ref{eq:Mcal}).
In the uncorrelated regime, adjacent $k_i$'s are all separated by a distance much larger than the entanglement length $\xi$,
whereas in the correlated regime, the distances between  adjacent $k_i$'s are smaller or on order of $\xi$.
For the uncorrelated regime, we have the following Lemma. 

\begin{Lemma}\label{lemma:uncorrelated_regime}
Suppose the adjacent sites with nonzero angle $\alpha_k$ are separated by distances much larger than entanglement range $\xi$.
Then 
\begin{flalign}
\left(\begin{array}{c}
\overline{T}^{(P)}_t(g_1)\\
\overline{T}^{(P)}_t(g_2)\\
:\\
\overline{T}^{(P)}_t(g_{|G|})\\
\end{array}\right)
=
\Pi_{k=1}^t \bra{\Psi} M_k(g_k,\alpha_k)\ket{\Psi}
\left(\begin{array}{c}
\overline{T}^{(P)}(g_1)\\
\overline{T}^{(P)}(g_2)\\
:\\
\overline{T}^{(P)}(g_{|G|})\\
\end{array}\right).
\label{eq:overline_T_t_P_factorize}
\end{flalign}
\end{Lemma}

{\em{Proof of Lemma~\ref{lemma:uncorrelated_regime}.}}
In the uncorrelated regime, by repeatedly using Lemma \ref{lemma:sting_factorize}, it can be seen that the expectation value in Eq. (\ref{eq:_product_M_rt}) factorizes and the expression can be further simplified to Eq. (\ref{eq:overline_T_t_P_factorize}). 
~~~$\Box$\medskip

We note that the factorization in Eq. (\ref{eq:overline_T_t_P_factorize}) does not hold in the correlated regime. 

In the uncorrelated regime, there is an alternative way of expressing the transformation in Eq. (\ref{eq:overline_T_t_P_factorize}) in terms of CPTP maps. 
We define logical gate $\overline{V}_k$ acting on the logical space $\mathcal{Q}$ as
\bea
\overline{V}_k:=\mathcal{P}e^{-i \frac{1}{2}\beta_k(\alpha_k) \overline{T}(g_k)}\mathcal{P}, \;\; -\pi \leq \alpha_k \leq \pi,\;g_k \in {\cal{G}}_k,
\eea
where according to Eq. (\ref{eq:Def_theta_k}), $\beta_k(\alpha_k)$ satisfies $\cos(\beta_k(\alpha_k))=\bra{\Psi} \cos\big[\mathcal{S}_k(g_k)\alpha_k\big]\ket{\Psi}$.
The CPTP map $\overline{\cal{V}}_k$ on logical space $\mathcal{Q}$ is defined as
\begin{equation}\label{eq:CPTP}
\overline{\cal{V}}_k:= \frac{1+\sigma_k(g_k,\alpha_k)}{2} [\overline{V}_k] + \frac{1-\sigma_k(g_k,\alpha_k)}{2} [\overline{V}^\dagger_k],
\end{equation}
in which the brackets $[\cdot]$ denote superoperators. 
%The rotation angle $\alpha_k$ and rotation axis specified by $g_k$ are implicit. 
For example, for a density matrix $\rho$ on $\mathcal{Q}$, we have
\begin{flalign}
\overline{\cal{V}}_k(\rho)=\frac{1+\sigma_k(g_k,\alpha_k)}{2}\overline{V}_k\rho\overline{V}_k^\dagger+\frac{1-\sigma_k(g_k,\alpha_k)}{2}\overline{V}^\dagger_k\rho\overline{V}_k.
\end{flalign}
On the other hand, suppose we work in the Heisenberg picture, then the action of $\overline{\cal{V}}_k$ on an operator $A$ is given by
\begin{flalign}
\overline{\cal{V}}^\dagger_k(A)=\frac{1+\sigma_k(g_k,\alpha_k)}{2}\overline{V}^\dagger_kA\overline{V}_k+\frac{1-\sigma_k(g_k,\alpha_k)}{2}\overline{V}_kA\overline{V}^\dagger_k.
\label{eq:CPTP_Heisenberg}
\end{flalign}

As before in Eq.~(\ref{eq:Vdef}), we define the concatenated operations $\overline{\cal{V}}_{\leq k}=\overline{\cal{V}}_{k}...\overline{\cal{V}}_{2}\overline{\cal{V}}_{1}$  (so that $\overline{\cal{V}}^\dagger_{\leq k}:=\overline{\cal{V}}^\dagger_1\overline{\cal{V}}^\dagger_2...\overline{\cal{V}}^\dagger_k$).
We have the following Lemma. 

\begin{Lemma}\label{lemma:CPTP}
In the uncorrelated regime, the transformation of projected evolved logical operators in Eq. (\ref{eq:overline_T_t_P_factorize}) can be equivalently written in the following form,
\bea
\overline{T}^{(P)}_t(g)=\overline{\cal{V}}^\dagger_{\leq t}\big(\overline{T}^{(P)}(g)\big),
\label{eq:Lemma_CPTP}
\eea
where $g\in G$.

\end{Lemma}

{\em{Proof of Lemma~\ref{lemma:CPTP}.}}
We will show that 
\bea
\mathcal{V}_k^\dagger(T^{(P)}(g))= \mathcal{M}_k(g_k,\alpha_k)(T^{(P)}(g)),
\eea
then Eq. (\ref{eq:Lemma_CPTP}) follows from induction. 
Using the definition of the matrix $M_k(g_k,\alpha_k)$, it is enough to show that 
\bea
e^{i \frac{1}{2}\beta_k(\alpha_k)\overline{T}(g_k)}\overline{T}(g)e^{-i \frac{1}{2}\beta_k(\alpha_k)\overline{T}(g_k)}=\overline{T}(g)
\label{eq:CPTP_proof_commu}
\eea
if $\kappa(g_k,g)=0$,
and 
\begin{flalign}
&e^{i \frac{1}{2}\beta_k(\alpha_k)\overline{T}(g_k)}\overline{T}(g)e^{-i \frac{1}{2}\beta_k(\alpha_k)\overline{T}(g_k)}=\nn\\
&\cos(\mathcal{S}_k\alpha_k) \overline{T}(g)+\sin(\beta_k(\alpha_k)) R_k(g,\alpha_k) i \overline{T}(g_k)\overline{T}(g)
\label{eq:CPTP_proof_acommu}
\end{flalign}
if $\kappa(g_k,g)=1$. 

Eq. (\ref{eq:CPTP_proof_commu}) is straightforward to see using $[\overline{T}(g_k),\overline{T}(g)]=0$ when $\kappa(g_k,g)=0$.

When $\kappa(g_k,g)=1$, for Eq. (\ref{eq:CPTP_proof_acommu}), we note that
\begin{flalign}
&e^{i \frac{1}{2}\beta_k(\alpha_k)\overline{T}(g_k)}\overline{T}(g)e^{-i \frac{1}{2}\beta_k(\alpha_k)\overline{T}(g_k)}\nn\\
&=e^{i \frac{1}{2}\beta_k(\alpha_k)Ad(\overline{T}(g_k))} \overline{T}(g),
\label{eq:TT_ad}
\end{flalign}
in which for an operator $D$, $Ad(\overline{T}(g_k))(D)=[\overline{T}(g_k),D]$. 
Using the anti-commutation relation between $\overline{T}(g_k)$ and $\overline{T}(g)$, and the property $[\overline{T}(g_k)]^2=1$ (which uses the fact that $v_{L,N+1}$ can be chosen such that $[v_{L,N+1}(g)]^2=1$, see Lemma 4 in Ref. \onlinecite{Raussendorf2023} for a proof), we obtain
\bea
\big[\frac{i}{2}Ad(\overline{T}(g_k))\big]^{2n}(\overline{T}(g))&=&(-)^n \overline{T}(g) \nn\\
\big[\frac{i}{2}Ad(\overline{T}(g_k))\big]^{2n+1}(\overline{T}(g))&=&(-)^n i\overline{T}(g_k)\overline{T}(g).
\label{eq:even_odd_TT}
\eea
Plugging Eq. (\ref{eq:even_odd_TT}) in Eq. (\ref{eq:TT_ad}), we obtain Eq. (\ref{eq:CPTP_proof_acommu}). ~~~$\Box$\medskip

We note that in the correlated regime, Eq. (\ref{eq:overline_T_t_P}) can also be written as a CPTP map,
though the expression is much more complicated.

%------------------------------------------------------------------------------------------------------------------------
\subsubsection{Approaching unitarity}

In the uncorrelated regime, each $M_k(g_k,\alpha_k)$ in Eq. (\ref{eq:overline_T_t_P_factorize}) is not a unitary matrix, or equivalently, the CPTP map $\overline{\mathcal{V}}_k$ in Eq. (\ref{eq:CPTP_Heisenberg}) is not a unitary transformation. 
However, as discussed in Ref. \onlinecite{Raussendorf2023}, by splitting a rotation angle into many pieces, unitarity can be approximated to arbitrarily good extent.  
More precisely, suppose an angle $\alpha$ is split into $n$ pieces, each with a small angle $\alpha/n$. 
The rotation sites are assumed to be mutually separated by distances much larger than the entanglement length. 
It has been shown in Ref. \onlinecite{Raussendorf2023} that the operator norm of the difference between $M_k(g_,\alpha/n)$ and $e^{i\frac{1}{2}\overline{T}(g)\nu_k(g)\alpha/n}$  is on order of $1/n^2$. 
By accumulating the $n$ gates with small angles, the difference between  $\Pi_{k=1}^n \bra{\Psi}M_k(g,\alpha_k)\ket{\Psi}$ ($\alpha_k=\alpha/n$) and $e^{i\frac{1}{2}\overline{T}(g)\nu_k(g)\alpha}$  is on order of $n\times1/n^2=1/n$, which approaches zero as $n\rightarrow \infty$. 

In the correlated regime, it has been shown in Ref. \onlinecite{Adhikary2023} that the same scaling applies.
Namely, even though the factorization in $\bra{\Psi}\Pi_kM_k(g,\alpha)\ket{\Psi}$ no longer applies, the difference between $\bra{\Psi}\Pi_kM_k(g,\alpha)\ket{\Psi}$ and $e^{i\frac{1}{2}\overline{T}(g)\alpha^\prime}$ is still on order of $1/n$ in the large $n$ limit, where $\alpha^\prime$ is related to $\alpha$ via a more complicated relation, which involves an average over string order parameters on both short and long length scales. 

We note that both the proof in Ref. \onlinecite{Raussendorf2023} for the uncorrelated regime and the proof in Ref. \onlinecite{Adhikary2023} for the correlated regime only involve the short-range entangled property of the state $\ket{\Psi}$ and does not rely on the assumptions of representations in Sec. \ref{sec:representation} (and Sec. 5.1 in Ref. \onlinecite{Raussendorf2023}). 
Therefore, they are applicable to the present formalism as well. 

%------------------------------------------------------------------------------------------------------------------------
\subsection{Block-local measurements} 
\label{subsec:block_measure}

For $h\in G$, the observable $T_N(h)$ can be measured in a block-local manner similar as Ref. \onlinecite{Raussendorf2023}, which we briefly discuss here. 
We describe the procedure in an inductive manner. 

Denote $s_i(g)$ to be the measurement outcome of the operator $O_i(g)$ located at block $i$, corresponding to the group element $g$ in $G$ (i.e., $(-)^{s_i(g)}$ is an eigenvalue of $O_i(g)$).
$O_i(g)$ will be constructed in an inductive manner. 

The induction begins at block $0$. 
Since $u_0$ is a linear representation, $u_0(g)$'s for all $g\in G$ commute and can be simultaneously measured.
$O_i(g)$ is chosen to be just $u_0(g)$. 
Denote $s_0(g)$ to be the measurement outcome of $O_0(g)$. 

Suppose all blocks $i$ ($0\leq i\leq k$ where $k\leq N$) have been measured. 
Define $q_k(g)$ as
\bea
q_k(g)=\sum_{i=1}^k s_i(g).
\label{eq:define_q_k_g}
\eea
Then $O_{k+1} (g)$ is constructed as
\bea
O_{k+1} (g)&=&e^{i \frac{1}{2}(-)^{q_k(g_{k+1})} \mathcal{S}_{k+1}(g_{k+1})\alpha_{k+1}} u_{k+1}(g)\nn\\
&&\cdot e^{-i\frac{1}{2} (-)^{q_k(g_{k+1})} \mathcal{S}_{k+1}(g_{k+1})\alpha_{k+1}},
\label{eq:Ok}
\eea
in which $q_k(g_{k+1})$ is known by induction. 
Clearly, $O_{k+1} (g)$'s are localized at block $k$ and commute for different $g\in G$, hence can be simultaneously measured.
The measured value of $O_{k+1} (g)$ is denoted as $(-)^{s_{k+1}(g)}$.

For the last step, $v_{L,N+1}$ at block $N+1$ is measured, with measurement result $(-)^{s_{N+1}(h)}$.  

After completing the measurements from blocks $0$ to $N+1$, the measured value of $T_N(h)$ in this round of measurement is given by $(-)^{\sum_{i=0}^{N+1} s_i(h)}$ (for an explanation, see Appendix \ref{app:measure_TNh}). 
The expectation value of $T_N(h)$ after many rounds of measurement is predicted to be given by Eq. (\ref{eq:overline_T_t_Psi}).

%------------------------------------------------------------------------------------------------------------------------
\subsection{Theorem for SPT-MBQC} 

Combining the above results together, we arrive at the following theorem for SPT-MBQC in 1D qudit systems. 

\begin{Theorem}\label{theorem:SPTMBQC}
Let $\ket{\Psi}$ be a many-body state of a 1D qudit system, symmetric under the group $G=(\mathbb{Z}_2)^m$. 
Then in the uncorrelated regime, the block-local measurements of the operator $\overline{T}(h)$ ($h\in G$) on $\ket{\Psi}$  simulates the action of the sequence of CPTP maps $\overline{\mathcal{V}}_{N}...\overline{\mathcal{V}}_1$ on $\ket{\Psi}\bra{\Psi}$. 

\end{Theorem}

{\em{Proof of Theorem~\ref{theorem:SPTMBQC}.}}
The measurement of $\overline{T}(h)$ on the state $\overline{\mathcal{V}}_{N}...\overline{\mathcal{V}}_1\ket{\Psi}$
is equivalent to the measurement of $\overline{T}_N(h)$ on $\ket{\Psi}$,
which can be performed in a block-local manner as discussed in Sec. \ref{subsec:block_measure}.
It has been proved that the expectation value $\bra{\Psi}\overline{T}_N(h)\ket{\Psi}$ obtained from the measurements is given by Eq. (\ref{eq:overline_T_t_Psi}),
which can be recast in terms of CPTP maps as shown in Lemma \ref{lemma:CPTP}.  ~~~$\Box$\medskip

Notice that in the small $\alpha_k$ limit, the CPTP map $\overline{\mathcal{V}}_k$ in Eq. (\ref{eq:CPTP}) is approximately a unitary transformation with rotation angle $\nu_k(g_k)\alpha_k$ where $\nu_k(g_k)$ is the bulk-to-end string order parameter defined in Eq. (\ref{eq:string_nu}).
Hence, the rotation angle implemented in MBQC is renormalized by a factor of $\nu_k(g_k)$ compared with the angle for tilting the measurement basis.
As long as $\nu_k(g_k)$ is nonzero, the implementation of MBQC is possible, though the efficiency is lowered when the value of $\nu_k(g_k)$ decreases. 

We not that Theorem \ref{theorem:SPTMBQC} can be generalized to the correlated regime as well,
except that the CPTP map must be determined from Eq. (\ref{eq:overline_T_t_P}) rather than Eq. (\ref{eq:overline_T_t_P_factorize}),
which no longer has a simple form as a composition of independent transformations.

%%%%%%%%%%%%%%%%%%%%%%%%%%%%%%%%%%%%%%%%%%%%%%%%%%%%%%
\section{Applications} 
\label{sec:applications}

We discuss two examples of the applications in Sec. \ref{sec:algebraic_MBQC},
including the spin-$1$ Haldane phase and the formalism in Ref. \onlinecite{Raussendorf2023}.

%------------------------------------------------------------------------------------------------------------------------
\subsection{SPT-MBQC in spin-$1$ Haldane phase} 

For the spin-$1$ chain in the Haldane phase,
we choose $\mathcal{G}_k\equiv G$, and define  $u_k(g)$, $\mathcal{S}_k(g)$ ($k=1,...,N$) as
\bea
u_k(R(\hat{\alpha},\pi))&=&e^{i\pi S_k^\alpha}\nn\\
\mathcal{S}_k(R(\hat{\alpha},\pi))&=&S_k^\alpha,
\eea
in which $\alpha=x,y,z$.
The projective representations at sites $0$ and $N+1$ are chosen as
\bea
v_{R,0}(\hat{\alpha},\pi)=v_{L,N+1}(\hat{\alpha},\pi)=\sigma^\alpha.
\eea
The linear representation $u_0$ can be chosen as
\bea
u_0(R(\hat{x},\pi))&=&X\nn\\
u_0(R(\hat{z},\pi))&=&I. 
\eea
As a result, $L_k(g)$ and $R_k(g,\alpha_k)$ can be determined as
\begin{flalign}
&L_k(R(\hat{\alpha},\pi))=u_0(R(\hat{\alpha},\pi))(\Pi_{j=1}^{k-1} e^{i\pi S_k^\alpha}) S_k^\alpha \nn\\
&R_k(R(\hat{\alpha},\pi),\alpha_k)=\frac{\sin\left[S_k^\alpha\alpha_k\right]}{\sin(\theta_k)} e^{i\pi S_k^\alpha} \left(\bigotimes_{j=k+1}^{N} e^{i\pi S_j^\alpha} \right) \sigma_{N+1}^\alpha,
\end{flalign}
where $\sigma_{N+1}^\alpha$ ($\alpha=x,y,z$) is the Pauli matrix at site $N+1$.
Notice that in the limit $\alpha_k\ll 1$ and keeping only up to linear order terms of $\alpha_k$,
$R_k(R(\hat{\alpha},\pi),\alpha_k)$ reduces to the bulk-to-end string order parameter in Eq. (\ref{eq:string_bulk_end}).

Applying Theorem \ref{theorem:SPTMBQC}, MBQC can be performed in the Haldane phase in spin-$1$ chains.
In fact, it is straightforward to see that the theorem applies to spin chains with spin value equal to arbitrary integer value in the SPT phase protected by $\mathbb{Z}_2\times \mathbb{Z}_2$ symmetry. 

%------------------------------------------------------------------------------------------------------------------------
\subsection{Relations to Ref. \onlinecite{Raussendorf2023}} 

We make some comments on the relation between the representations in Sec. \ref{sec:representation} and those in Sec. 5.1.1 in Ref. \onlinecite{Raussendorf2023}.
The representations at boundaries in Sec. \ref{sec:representation} are the same as those in Sec. 5.1.1 in Ref. \onlinecite{Raussendorf2023}.
The only difference is in the bulk.
In the present formalism, there is no definition of projective representation $v_{L,i}$ and $v_{R,i}$ ($1\leq i\leq n$)  in the bulk. 
The conditions are loosened compared with Ref. \onlinecite{Raussendorf2023}.

In fact, if we define $\mathcal{S}_i=v_{R,i}$, then the conditions in Ref. \onlinecite{Raussendorf2023} imply the conditions in Sec. \ref{subsec:conditions}. 
Or more precisely, Eq. (\ref{eq:condition_Sg}) holds.  
This is because  $\forall g\in G,g^\prime \in \mathcal{G}_i$,
\begin{flalign}
&u_i(g) v_{R,i}(g^\prime) [u_i(g)]^{-1}\nn\\
&=v_{L,i}(g)v_{R,i}(g)v_{R,i}(g^\prime)[v_{R,i}(g)]^{-1}[v_{L,i}(g)]^{-1}\nn\\
&=(-)^{\kappa(g,g^\prime)}v_{L,i}(g)v_{R,i}(g^\prime)[v_{L,i}(g)]^{-1}\nn\\
&= (-)^{\kappa(g,g^\prime)}v_{R,i}(g),
\end{flalign}
in which
in the second equality, we have used (see Eq. (26) in Ref. \onlinecite{Raussendorf2023})
\begin{flalign}
&\forall g\in G,g^\prime \in \mathcal{G}_i,\nn\\
&v_{R,i}(g) v_{R,i}(g') = (-1)^{\kappa(g,g')} v_{R,i}(g') v_{R,i}(g),  
\label{eq:proj_L}
\end{flalign}
and in the third equality, we have used  (see Eq. (27) in Ref. \onlinecite{Raussendorf2023})
\bea
[v_{L,i}(g),v_{R,i}(g^\prime)]=0, \forall g\in G,g^\prime\in\mathcal{G}_i.
\label{eq:v_L_R_commute}
\eea 

%%%%%%%%%%%%%%%%%%%%%%%%%%%%%%%%%%%%%%%%%%%%%%%%%%%%%%
\section{Summary}
\label{sec:summary}

We have formulated an algebraic framework for measurement-based quantum computation in one-dimensional symmetry protected topological states, which applies to both half-odd-integer and integer spin systems,
generalizing the framework in Ref. \onlinecite{Raussendorf2023}.
The technical advance we have made to achieve this is to circumvent the need of projective representations in the bulk. 
The advantage of this generalized framework is that systems in the same symmetry protected topological phase can be treated in the same setting regardless of their spin values. 
The computational order parameter is identified, which quantitatively characterizes the efficiency of measurement-based quantum computation. 
For the  Haldane phase of spin-$1$ chains, this computational order parameter coincides with the conventional string order parameter.

%%%%%%%%%%%%%%%%%%%%%%%%%%%%%%%%%%%%%%%%%%%
{\it Acknowledgments}
WY is supported by the startup funding in Nankai University. 
RR is funded by the Humboldt foundation. 
AA and RR were funded by USARO (W911NF2010013). 
WY, AA and RR were funded from the Canada First Research Excellence Fund, Quantum Materials and Future Technologies Program. 
WY and RR were funded by NSERC, through the European-Canadian joint project FoQaCiA (funding reference number 569582-2021).

%%%%%%%%%%%%%%%%%%%%%%%%%%%%%%%%%%%%%%%%%%%%%%%%%%%%%%%%%%%%%%
%%%%%%%%%%%%%%%%%%%%%%%%%%%%%%%%%%%%%%%%%%%%%%%%%%%%%%%%%%%%%%
\appendix

%%%%%%%%%%%%%%%%%%%%%%%%%%%%%%%%%%%%%%%%%%%%%%%%%%%%%%%%%%%%%%
\section{Measuring $T_N(h)$}
\label{app:measure_TNh}

Define $\tilde{u}_0(g)$ to be $u_0(g)$ ($g\in G$).
Define $\tilde{u}_k(g)$ in an inductive way ($1\leq k\leq N$) as
\bea
\tilde{u}_k(g)=\tilde{V}^\dagger_k u_k(g) \tilde{V}_k,
\label{eq:u_tilde_alter}
\eea
in which $\tilde{V}_k$ is determined by $\tilde{u}_j(g_k)$ ($j\leq k-1$) as 
\bea
\tilde{V}_k=e^{i\frac{1}{2} \big(\Pi_{j=1}^{k-1} \tilde{u}_{j}(g_k)\big) \mathcal{S}_k(g_k)\alpha_k }. 
\label{eq:V_tilde_alter}
\eea

\begin{Lemma}\label{lemma:commute_ucal}
 $[\tilde{u}_i(g_i),\tilde{u}_j(g_j)]=0$, $\forall 1\leq i,j\leq N$.
\end{Lemma}

{\em{Proof of Lemma~\ref{lemma:commute_ucal}.}}
We prove by induction that $[\tilde{u}_i(g),\tilde{u}_j(g^\prime)]=0$, $\forall 1\leq i,j\leq n$, $\forall g,g^\prime\in G$, where $n\leq N$.

When $n=0$, $[u_0(g),u_0(g^\prime)]=0$ since $u_0$ is a linear representation. 

Suppose the property holds for $n$. 
We want to show that it is also valid for $n+1$.
It clear that $[\tilde{u}_{n+1}(g),\tilde{u}_{n+1}(g^\prime)]=\tilde{V}_{n+1}^\dagger [u_{n+1}(g),u_{n+1}(g^\prime)] \tilde{V}_{n+1}=0$.
Then it is enough to show that $[\tilde{u}_{n+1}(g),\tilde{u}_j(g^\prime)]=0$ for $j<n+1$.
Since $\tilde{u}_j(g^\prime)$ only has nonvanishing support on the blocks $\{0,1,2,...,j\}$, it commutes with $\mathcal{S}_{n+1}(g_{n+1})$ and $u_{n+1}(g)$.
By induction hypothesis and $[\tilde{u}_j(g^\prime),\mathcal{S}_{n+1}(g_{n+1})]=0$, we have $[\tilde{u}_j(g^\prime),\tilde{V}_{n+1}]=0$.
Then since $[\tilde{u}_j(g^\prime),u_{n+1}(g)]=0$ and $[\tilde{u}_j(g^\prime),\tilde{V}_{n+1}]=0$, we obtain $[\tilde{u}_j(g^\prime),\tilde{u}_{n+1}(g)]=0$.~~~$\Box$\medskip

\begin{Lemma}\label{lemma:commute_V_ucal}
$\forall g_k\in \mathcal{G}_k, g\in G$, 
\begin{enumerate}
\item{ $[\tilde{V}_k(g_k),u_j(g)]=0$, when $j>k$,
\label{lemma_commute_V_ucal_1}
}
\item{ $[\tilde{V}_k(g_k),\tilde{u}_j(g)]=0$, when $j<k$.
\label{lemma_commute_V_ucal_2}
}
\end{enumerate}
\end{Lemma}

{\em{Proof of Lemma~\ref{lemma:commute_V_ucal}.}}
When $j>k$, the support of $\tilde{V}_k(g_k)$ does not contain $j$, since its support is on the blocks $i$ where $i$ is smaller than or equal to $k$. Hence $[\tilde{V}_k(g_k),u_j(g)]=0$.  

When $j<k$, $[\tilde{V}_k(g_k),\tilde{u}_j(g)]=0$ follows from $[\tilde{u}_j(g),\mathcal{S}_k(g_k)]=0$ and Lemma \ref{lemma:commute_ucal}.
~~~$\Box$\medskip

\begin{Lemma}\label{lemma:Vtilde_alternative}
$\tilde{V}_k$ can be written as
\bea
\tilde{V}_k=\tilde{V}^\dagger_{k-1}...\tilde{V}^\dagger_1  V_k \tilde{V}_1 ...\tilde{V}_{k-1}.
\eea
\end{Lemma}

{\em{Proof of Lemma~\ref{lemma:Vtilde_alternative}.}}
For $j< k$, from Eq. (\ref{lemma_commute_V_ucal_1}) in Lemma \ref{lemma:commute_V_ucal}, we have
\bea
\tilde{V}_j^\dagger \tilde{V}_{j-1}^\dagger...\tilde{V}_1^\dagger u_{j}(g)\tilde{V}_1 ... \tilde{V}_{j-1} \tilde{V}_j
&=&\tilde{V}_j^\dagger u_{j}(g)\tilde{V}_j\nn\\
&=&\tilde{u}_j(g).
\eea
Then using Eq. (\ref{lemma_commute_V_ucal_2}) in Lemma \ref{lemma:commute_V_ucal}, we obtain (for $j<k$)
\bea
\tilde{u}_j(g)=\tilde{V}^\dagger_{k-1}...\tilde{V}^\dagger_1  u_j(g) \tilde{V}_1 ...\tilde{V}_{k-1}
\eea
Hence $\tilde{V}_k$ in Eq. (\ref{eq:V_tilde_alter}) can be written as
\bea
\tilde{V}_k&=&e^{i\frac{1}{2} \big(\Pi_{j=1}^{k-1} \tilde{V}^\dagger_{k-1}...\tilde{V}^\dagger_1 u_{j}(g_k) \tilde{V}_1 ...\tilde{V}_k\big) \mathcal{S}_k(g_k)\alpha_k }\nn\\
&=&\tilde{V}^\dagger_{k-1}...\tilde{V}^\dagger_1 V_k \tilde{V}_1 ...\tilde{V}_{k-1},
\eea
completing the proof. ~~~$\Box$\medskip

\begin{Lemma}\label{lemma:Vtilde_product}
$V_k...V_1=\tilde{V}_1...\tilde{V}_k$, $1\leq k\leq N$.
\end{Lemma}

{\em{Proof of Lemma~\ref{lemma:Vtilde_product}.}}
Using Lemma \ref{lemma:Vtilde_alternative}, we obtain
\bea
\tilde{V}_1...\tilde{V}_k&=&V_k(\tilde{V}_1...\tilde{V}_{k-1})\nn\\
&=&V_k(V_{k-1}\tilde{V}_1...\tilde{V}_{k-2})\nn\\
&=&...\nn\\
&=& V_k...V_1.  ~~~\Box
\eea \medskip

\begin{Lemma}\label{lemma:TNh_alternative}
$\overline{T}_N(h)=\big[\Pi_{k=1}^N \tilde{u}_k(h) \big]v_{L,N+1}(h)$.
\end{Lemma}

{\em{Proof of Lemma~\ref{lemma:TNh_alternative}.}}
Using Eq. (\ref{eq:GateSeqs}) and Lemma \ref{lemma:Vtilde_product}, we obtain
\begin{flalign}
\overline{T}_N(h)= \tilde{V}_N^\dagger ... \tilde{V}_1^\dagger\big[ \otimes_{i=0}^N u_i(h)\otimes v_{L,N+1}(h)\big] \tilde{V}_1...\tilde{V}_N.
\label{eq:TNh_alternative_0}
\end{flalign}
By Eq. (\ref{eq:u_tilde_alter}) and Lemma \ref{lemma:commute_V_ucal}, it can be seen that
\bea
\tilde{V}_N^\dagger ... \tilde{V}_1^\dagger u_i(h)  \tilde{V}_1 ... \tilde{V}_N=\tilde{u}_i(h).
\label{eq:TNh_alternative_u}
\eea
Since $\tilde{V}_k$ ($1\leq k\leq N$) has no support on block $N+1$, we have 
\bea
\tilde{V}_N^\dagger ... \tilde{V}_1^\dagger v_{L,N+1}  \tilde{V}_1 ... \tilde{V}_N=v_{L,N+1}.
\label{eq:TNh_alternative_v}
\eea
The proof is completed by plugging Eq. (\ref{eq:TNh_alternative_u}) and Eq. (\ref{eq:TNh_alternative_v}) in Eq. (\ref{eq:TNh_alternative_0}). ~~~$\Box$\medskip

Now we are prepared to show that the procedure described in Sec. \ref{subsec:block_measure} measures $\overline{T}(h)$ in a block-local way.
Define $\ket{s_k, O_k}_k$ to be the basis state at block $k$ corresponding to eigenvalues $(-)^{s_k(g)}$ of the operators $O_k(g)$ ($g\in G$), where $O_k(g)$ is defined in Eq. (\ref{eq:Ok}).
Define $\ket{s_{N+1}}_{N+1}$ to be the basis state at block $N+1$ corresponding to eigenvalue $(-)^{s_{N+1}(h)}$ of the operator $s_{N+1}(h)$.
Define $\ket{\vec{s}}$ as
\bea
\ket{\vec{s}}=\big(\otimes_{k=0}^{N} \ket{s_k,O_k}_k\big)\otimes \ket{s_{N+1}}_{N+1}.
\eea

In fact, the action of $\tilde{u}_k(g)$ ($1\leq k\leq N$) on $\ket{\vec{s}}$ is given by
\bea
\tilde{u}_k(g)\ket{\vec{s}}=(-)^{s_k(g)} \ket{\vec{s}}. 
\label{eq:u_ket_vec_s}
\eea
We show this by induction.
For $k=0$, it can be clearly seen that acting $\tilde{u}_0(g)=u_0(g)$ on $\ket{\vec{s}}$ gives $s_0(g)$.
Suppose Eq. (\ref{eq:u_ket_vec_s}) holds for $k\leq n$. 
Then for $k=n+1$, according to Eq. (\ref{eq:V_tilde_alter}), we know that 
\bea
\tilde{V}_{n+1}=e^{i\frac{1}{2} q_n(g)\mathcal{S}_{n+1}(g_{n+1})\alpha_{n+1}}, 
\label{eq:V_nplus1}
\eea
where $q_n(g)$ is defined in Eq. (\ref{eq:define_q_k_g}). 
Plugging Eq. (\ref{eq:V_nplus1}) into Eq. (\ref{eq:u_tilde_alter}) and comparing with Eq. (\ref{eq:Ok}), we obtain
\bea
\tilde{u}_{n+1}(g)\ket{\vec{s}} = O_{k+1}(g)\ket{\vec{s}}=(-)^{s_{n+1}(g)}\ket{\vec{s}}.
\eea

Using Eq. (\ref{eq:u_ket_vec_s}) and $v_{L,N+1}(h)\ket{s_{N+1}}_{N+1}=(-)^{s_{N+1}(h)}\ket{s_{N+1}}_{N+1}$, 
we obtain
\bea
\overline{T}(h)\ket{\vec{s}}=(-)^{\sum_{j=0}^{N+1} s_j(h) }\ket{\vec{s}}.
\eea
Namely, $\ket{\vec{s}}$ is an eigenvector of $\overline{T}(h)$ with eigenvalue $(-)^{\sum_{j=1}^{N+1} s_j(h)}$.
Therefore, the procedure described in Sec. \ref{subsec:block_measure} is a block-local and projective measurement for $\overline{T}(h)$.

%%%%%%%%%%%%%%%%%%%%%%%%%%%%%%%%%%%%%%%%%%%%%%%%%%%%%%%%%%%%%%

\end{document}